\documentstyle[12pt,epsfig]{article}
\newcommand{\ssim}{\scriptstyle \sim}
\begin{document}
\setlength{\parskip}{0.4cm}
\setlength{\baselineskip}{0.66cm}
\begin{titlepage}
\setlength{\parskip}{0.25cm}
\setlength{\baselineskip}{0.25cm}
\begin{flushright}
{\tt hep-ph/9806404} \hfill DO-TH 98/07 \\[1mm]  
WUE-ITP-98-019\\[2mm]
June 1998 
\end{flushright}
\vspace{1.5cm}
\begin{center}
\LARGE
{\bf Dynamical Parton Distributions Revisited}
\vspace{1.5cm}

\large
M. Gl\"uck$^1$, E.\ Reya$^1$ and A. Vogt$^2$\\
\vspace{1.0cm}

\normalsize
$^1${\it Institut f\"{u}r Physik, Universit\"{a}t Dortmund}\\ 
{\it D-44221 Dortmund, Germany} \\
\vspace{0.4cm}

$^2${\it Institut f\"ur Theoretische Physik, Universit\"at W\"urzburg} \\
{\it Am Hubland, D-97074 W\"urzburg, Germany}

\vspace{2.5cm}
\end{center}
\begin{abstract}
Dynamical parton densities, generated radiatively from valence--like
inputs at some low resolution scale, are confronted with recent small--$x$
data on deep inelastic and other hard scattering processes.  It is shown
that within theoretical uncertainties our previous (1994) dynamical/radiative
parton distributions are compatible with most recent data and still applicable
within the restricted accuracy margins of the presently available 
next--to--leading order calculations.  Due to recent high precision
measurements we also present an updated, more accurate, version of our 
(valence--like) dynamical input distributions.  Furthermore, our
perturbatively stable parameter--free dynamical predictions are extended
to the extremely small--$x$ region,  $10^{-8}$ \raisebox{-0.1cm}{$\stackrel{<}
{\ssim}$} $x$ \raisebox{-0.1cm}{$\stackrel{<}{\ssim}$} $10^{-5}$, relevant 
to questions concerning ultra--high--energy \mbox{cosmic} ray and neutrino 
astronomy.
\end{abstract}
\end{titlepage}
 
\section{Introduction}
\vspace*{-2mm}

The guiding physical idea underlying the dynamical (radiative) parton model is 
that the steep behavior of the momentum distributions $xf(x,Q^2)\,\, (f=q,\,
\bar{q},\, g)$ at $x$ \raisebox{-0.1cm}{$\stackrel{<}{\ssim}$} $10^{-2},\,\,$ 
\mbox{$Q^2$ \raisebox{-0.1cm}{$\stackrel{>}{\ssim}$} 1 GeV$^2$} is a purely 
perturbative phenomenon.  In fact, in \cite{ref1,ref2,ref3} {\it non}--steep 
(valence--like) initial distributions $xf(x,\mu^2)$ at some low scale 
$\mu\approx 0.6$ GeV were suggested in order to predict, purely dynamically,
the rise of $x\bar{q}$ and $xg$ in the above range of $x$ and $Q^2$.  This 
prediction was subsequently confirmed by the measurements of $F_2^p(x,Q^2)$ and 
$xg(x,Q^2)$ at HERA \cite{ref4,ref5}.  As was stated in \cite{ref2,ref3}, the 
available {\it pre}--HERA data at $x>10^{-2}$ \mbox{utilized to fix} the 
valence--like input distributions still allowed for a slight, typically 
O(10\%), variation (increase) of $\mu$ which did not affect $F_2(x\! >\! 
10^{-2},\, Q^2)$ but resulted in an about 10\% (20\%) uncertainty of the 
radiative predictions at $x=10^{-3}\, (10^{-4})$.  Hence the O(10\%) 
discrepancies between the distributions of \cite{ref3} and recent precision 
measurements at HERA \cite{ref6,ref7,ref8} are obviously neither unexpected, 
nor do they invalidate the notion of a radiative, i.e.\ dynamical, origin for 
the steep rise at $x$ \raisebox{-0.1cm}{$\stackrel{<}{\ssim}$} $10^{-2}$ and 
$Q^2$ \raisebox{-0.1cm}{$\stackrel{>}{\ssim}$} $1$ GeV$^2$.  Indeed, a 
fine--tuning of $\mu$ and/or $f(x,\mu^2)$ was always understood 
\cite{ref2,ref3} to be necessary in due course.

Given the accuracy of recent HERA small--$x$ data, as well as new large--$x$ 
constraints, mainly on the flavor decomposition of the quark sea, it is now 
appropriate to perform an update of our previous dynamical parton distributions 
and to follow the effects of these fine--tunings on the predictions for 
$F_2(x,Q^2)$, $g(x,Q^2)$ and other relevant deep inelastic observables.

Section 2 will be devoted mainly to a discussion concerning the small--$x$
implications of the recent H1 and ZEUS high precision data 
\cite{ref6,ref7,ref8} on $F_2^p(x,Q^2)$ and $g(x,Q^2)$ and how they (slightly)
modify our previous \cite{ref3} valence--like LO/NLO input gluon and sea
distributions.  Further data on $F_2^c(x,Q^2)$ and fixed--target results on 
$F_2^p,\, F_2^n/F_2^p$ and $d_v/u_v$, as well as data on $pp(d)\to \mu^+\mu^-X$ 
and $p\bar{p}\to W^{\pm}X$ asymmetries relevant for fixing and testing our 
$g,\, u_v,\, d_v,\, \bar{d}-\bar{u}$ and $u/d$ densities are presented and 
compared with the present update and with our previous GRV(94) results in 
Section 3.  Furthermore, we also extend here our perturbatively stable and 
parameter--free dynamical small--$x$ predictions to the ultra small--$x$ 
region ($10^{-8}$ \raisebox{-0.1cm}{$\stackrel{<}{\ssim}$} $x$ 
\raisebox{-0.1cm}{$\stackrel{<} {\ssim}$} $10^{-5}$), relevant to questions
concerning ultra--high--energy cosmic ray and neutrino astronomy.  Finally, in 
Section 4 we present a brief summary and a general discussion concerning the 
status of the dynamical (radiative) parton model in the light of present and 
future data and its application in theoretical NLO analyses. 
 
\section{Consequences of recent data on {\boldmath $F_2(x,Q^2)$} and 
 \mbox{parton} densities}

As stated in the Introduction we intend to study the implications of 
modifications of our original GRV(94) input \cite{ref3}. We shall proceed in a 
stepwise manner and begin with a mere modification of the input scale $\mu^2$ 
keeping everything else unchanged.  The results shown in Figs.\ 1 and 2 
illustrate the point that the previously noted \cite{ref2,ref3} uncertainty in 
$\mu^2$ can accommodate the main discrepancies between \cite{ref3} and the 
recent data \cite{ref6,ref7,ref8} on $F_2(x,Q^2)$ {\it and\/} the 
experimentally extracted $xg(x,Q^2)$.  It should be noted that our present new 
(modified) analysis yields results which lie in between the two curves shown in 
Fig.\ 1 as will be discussed and illustrated in Fig.\ 3.  Our new results 
(solid curve) for $xg(x,Q^2)$ at $Q^2=20$ GeV$^2$ are shown as well in Fig.\ 2
where they practically coincide with the one of GRV(94) with $\mu_{\rm{NLO}}$
increased by 10\% (dotted curve).  The fact that the somewhat too steep GRV(94) 
predictions for $F_2(x,Q^2)$ and $xg(x,Q^2)$ at, say, $x=10^{-4}$ can 
{\it both\/} be corrected with the new slightly modified valence--like input 
constitutes a nontrivial confirmation of the radiative model.

Let us now turn to the update of our LO/NLO GRV(94) densities which consists of 
a fine--tuning of the valence--like input densities for $f(x,\mu^2)$ {\it as 
well as\/} of the input scale $\mu$.  This fine--tuning of the input results 
mainly from new HERA data \cite{ref6,ref7,ref8} on $F_2(x,Q^2)$ and to a 
certain extent also from new large--$x$ results (such as asymmetry measurements 
of Drell--Yan dilepton production in $pp$ and $pd$ collisions 
\cite{ref9,ref10}) discussed in more detail in Section 3.  For the running 
coupling $\alpha_s(Q^2)$ at the next--to--leading order (NLO) we have utilized 
the {\it exact\/} numerical solution\footnote{Alternatively, the exact solution 
can be written implicitly 
\begin{displaymath}
\ln\frac{Q^2}{\tilde{\Lambda}^2_{\overline{\rm MS}}} = \frac{4\pi}
{\beta_0\alpha_s(Q^2)}\,-\, \frac{\beta_1}{\beta_0^2} \, \ln
\left[ \frac{4\pi}{\beta_0\alpha_s(Q^2)}\, +\, \frac{\beta_1}{\beta_0^2}
\right] ,
\end{displaymath}
and our new NLO results correspond, for $\alpha_s(M_Z^2)=0.114$, to the
following values of $\tilde{\Lambda}^{(f)}_{\overline{\rm MS}}:\, \tilde
{\Lambda}^{(3,4,5,6)}_{\overline{\rm MS}}=299.4,\, 246,\, 167.7,\, 67.8$ MeV.} 
of
\begin{equation}
\frac{d\, \alpha_s(Q^2)}{d\, \ln(Q^2)}= -\frac{\beta_0}{4\pi}\, \alpha_s^2(Q^2)
\, -\, \frac{\beta_1}{16\pi^2}\, \alpha_s^3(Q^2)
\end{equation}
with $\beta_0=11-2f/3$ and $\beta_1=102-38f/3$, which is nowadays used for NLO 
analyses \cite{ref6,ref8,ref11,ref12,ref13} since it is stable against yet 
higher order contributions, thus being more appropriate in the low $Q^2$ 
region.  This is in contrast to using \cite{ref1,ref2,ref3} the approximate 
solution
\begin{equation}
\frac{\alpha_s(Q^2)}{4\pi} \simeq \frac{1}{\beta_0\, \ln(Q^2/\Lambda^2)}\, 
-\, \frac{\beta_1}{\beta_0^3} \,\, \frac{\ln \ln(Q^2/\Lambda^2)}
{[\ln(Q^2/\Lambda^2)]^2}
\end{equation}
\\[-3mm]
which is sufficiently accurate for $Q^2$ \raisebox{-0.1cm}{$\stackrel{>}
{\ssim}$} $m_c^2$.  Here, $\Lambda$ refers in NLO to $\Lambda\equiv
\Lambda_{\overline{\rm MS}}$ and in LO ($\beta_1=0$) to $\Lambda\equiv 
\Lambda_{\rm LO}$.  We have chosen $\alpha_s(M_Z^2)=0.114$ for obtaining our 
exact numerical NLO solutions from (1) for $Q^2\geq\mu_{\rm NLO}^2$.  This 
choice, which is slightly preferred in our present analysis, agrees with the 
average value of the space--like momentum--transfer measurements
\cite{ref13,ref14,ref15}, $\alpha_s(M_Z^2)=0.114\pm0.005$.  The statistically 
dominating time--like $e^+e^-$ LEP $Z^0$--data imply a somewhat larger `world 
average' \cite{ref13,ref14} of $\alpha_s(M_Z^2)= 0.118\pm 0.005$ with an error 
which is to some extent uncertain and debatable.  It should be kept in mind, 
however, that LEP data ($Z^0$ hadronic decays) allow \cite{ref16} also for a 
much smaller strong coupling, $\alpha_s(M_Z^2)=0.101\pm 0.013$.

Should significantly higher values of $\alpha_s$, e.g.~$\alpha_s(M_Z^2)=0.118$,
turn out to be undebated and everywhere unique, then our input scale $\mu$ will 
obviously increase closer to 1 GeV which compels one to give up the strict 
valence--like sea input $x\bar{q}(x,\mu^2)$, but {\it not\/} the valence--like 
gluon input $xg(x,\mu^2)$, in order to reproduce all $F_2^p(x,Q^2)$ data at 
$Q^2$ \raisebox{-0.1cm}{$\stackrel{>}{\ssim}\:$} \mbox{1 GeV$^2$} \cite{ref11}.
Alternatively one may keep the valence--like sea input $x\bar{q}(x,\mu^2)$, 
also for $\alpha_s(M_Z^2)= 0.118$, as a parameter--free seed for the small--$x$
structure of $F_2(x,Q^2>\mu^2)$, in case one intends to predict and explain the 
HERA--data merely above $Q^2\simeq 3$ GeV$^2$ \cite{ref12}.

Furthermore the conventional approximate formula~(2), being sufficiently 
reliable for $Q^2$ \raisebox{-0.1cm}{$\stackrel{>}{\ssim}$} $m_c^2$, 
corresponds in our case to $\Lambda_{\overline{\rm MS}}^{(4,5,6)}= 257,\, 
173.4,\, 68.1$ MeV which reproduces the exact solutions to even better than 
0.5\% for $Q^2\geq 5$ GeV$^2$.  Our LO results correspond to $\Lambda_{\rm LO}
^{(3,4,5,6)} = 204,\, 175,\, 132,\, 66.5$ MeV which leads to the (theoretically 
less relevant) value of $\alpha_s^{\rm LO}(M_Z^2) = 0.125$.  In both cases we 
have used for the $\alpha_s$ matchings
\begin{equation}
m_c=1.4\, {\rm GeV},\quad\quad m_b=4.5\, {\rm GeV},\quad\quad
m_t=175 \, {\rm GeV}.
\end{equation}
These masses are used in all our subsequent LO and NLO analyses for heavy quark 
production.  In particular the value of $m_c$, which is slightly lower than the 
one previously employed in \cite{ref1,ref2,ref3}, is favored.

The free parameters of the non--singlet input densities $u_v,\, d_v,\,
\Delta\equiv \bar{d}-\bar{u}$ and of the valence--like singlet input
distributions $\bar{d}+\bar{u}$ and $g$ at $Q^2=\mu^2$ have been fixed
using the following data sets:  the published 1994 and 1995 HERA $F_2^p$
results \cite{ref6,ref7} for \mbox{$Q^2\geq 2$ GeV$^2$}; the fixed target
$F_2^p$ data of SLAC \cite{ref17}, BCDMS \cite{ref18,ref19}, NMC \cite{ref20} 
and E665 \cite{ref21} subject to the standard cuts $Q^2\geq4$ GeV$^2$ and 
$ W^2=Q^2(\frac{1}{x}-1)+m_p^2\geq 10$ GeV$^2$; the structure function ratios 
$F_2^n/F_2^p$ \cite{ref22,ref23}, together with the $u_v/d_v$ results from 
$\stackrel{{\scriptscriptstyle (-)}}{\nu}\!\!p(d)$ DIS \cite{ref24}; 
the Drell-Yan muon--pair production data of E605 \cite{ref25} for 
$d^2\sigma^{pN}/dx_F dM_{\mu^+\mu^-}$ and of NA51 \cite{ref9} and E866 
\cite{ref10} for the cross section ration $\sigma^{pd}/\sigma^{pp}$.  The input 
fit parameters/normalizations of $u_v$ and $d_v$ are further constrained by 
$\int_0^1 u_v dx=2$ and $\int_0^1 d_v dx = 1$, and the ones of the gluon 
density by the energy--momentum conservation relation
\begin{equation}
\int_0^1 x\left[ u_v(x,\mu^2) + d_v(x,\mu^2) + 2 \bar{u}(x,\mu^2)
  + 2\bar{d}(x,\mu^2) + g(x,\mu^2) \right] \, dx = 1.
\end{equation}
The resulting LO input distributions at $Q^2=\mu_{\rm LO}^2=0.26$ GeV$^2$ are 
then given by
\begin{eqnarray}
xu_v(x,\mu_{\rm LO}^2) & = & 1.239\,\,x^{0.48}\,(1-x)^{2.72}\, 
    (1-1.8\sqrt{x} + 9.5x)\nonumber\\
xd_v(x,\mu_{\rm LO}^2) & = & 0.614\,\, (1-x)^{0.9}\,\, 
     xu_v(x,\mu_{\rm LO}^2)\nonumber\\
x\Delta(x,\mu_{\rm LO}^2) & = & 0.23 \,\, x^{0.48}\,(1-x)^{11.3}\,
    (1-12.0\sqrt{x} + 50.9x)\nonumber\\
x(\bar{u}+\bar{d})(x,\mu_{\rm LO}^2) & = & 1.52\,\, x^{0.15}\, (1-x)^{9.1}\,
    (1-3.6\sqrt{x} + 7.8x)\nonumber\\
xg(x,\mu_{\rm LO}^2) & = & 17.47\,\, x^{1.6}\, (1-x)^{3.8}\nonumber\\
xs(x,\mu_{\rm LO}^2) & = & x\bar{s}(x,\mu_{\rm LO}^2) = 0
\end{eqnarray}
where $\Delta\equiv\bar{d}-\bar{u}$.  It is interesting to note that
$\mu_{\rm LO}=2.5\,\Lambda_{\rm LO}^{(3)}=2.9\,\Lambda_{\rm LO}^{(4)}$
and $\alpha_s^{\rm LO}(\mu_{\rm LO}^2)/\pi = 0.24$.  The corresponding
NLO($\overline{\rm{MS}}$) input at $Q^2=\mu_{\rm NLO}^2=0.40$ GeV$^2$ 
is~\footnote{The power
$a_{\rm sea}$ of the valence--like LO/NLO sea input densities $x(\bar{u}+
\bar{d})(x,\mu^2)\sim x^{a_{\rm sea}}$, as $x\to 0$, depends strongly on the 
choice of $\mu$, i.e.~on the chosen value for $\alpha_s(M_Z^2)$ or, 
equivalently, on $\Lambda^{(3)}$.  We would obtain almost equally agreeable 
results if we continue to use the {\it same\/} $\alpha_s(Q^2)$ as in our 
previous analyses \cite{ref1,ref2,ref3} which was based on the approximate 
evolution formula (2).  Matching the approximate NLO(${\overline{\rm MS}})\,
\alpha_s(Q^2)$ to the exact numerical solution of eq.~(1) at, say, 
$Q^2=3$ GeV$^2$ results in $\alpha_s(M_Z^2)=0.110$.  As discussed above, this 
$\alpha_s$ value lies at the lower end of the presently allowed experimental
bounds.  In this case one obtains $\mu_{\rm NLO}^2\simeq 0.30$ GeV$^2$ and
$a_{\rm sea}^{\rm NLO} \simeq 0.35$, instead of 0.20 in eq.~(6).}  
\begin{eqnarray}
xu_v(x,\mu_{\rm NLO}^2) & = & 0.632\,\,x^{0.43}\,(1-x)^{3.09}\, 
    (1+18.2x)\nonumber\\
xd_v(x,\mu_{\rm NLO}^2) & = & 0.624\,\, (1-x)^{1.0}\,\, 
     xu_v(x,\mu_{\rm NLO}^2)\nonumber\\
x\Delta(x,\mu_{\rm NLO}^2) & = & 0.20 \,\, x^{0.43}\,(1-x)^{12.4}\,
    (1-13.3\sqrt{x} + 60.0x)\nonumber\\
x(\bar{u}+\bar{d})(x,\mu_{\rm NLO}^2) & = & 1.24\,\, x^{0.20}\, (1-x)^{8.5}\,
    (1-2.3\sqrt{x} + 5.7x)\nonumber\\
xg(x,\mu_{\rm NLO}^2) & = & 20.80\,\, x^{1.6}\, (1-x)^{4.1}\nonumber\\
xs(x,\mu_{\rm NLO}^2) & = & x\bar{s}(x,\mu_{\rm NLO}^2) = 0.
\end{eqnarray}
Note again that $\mu_{\rm NLO}=2.1\,\, \tilde{\Lambda}_{\overline{\rm MS}}^
{(3)}=2.6\,\,\tilde{\Lambda}_{\overline{\rm MS}}^{(4)}$ and $\alpha_s(\mu_{\rm 
NLO}^2)/\pi=0.18$, and that there is a correlation between the chosen value of 
$\alpha_s(M_Z^2)$ and the resulting value for $\mu_{\rm NLO}$, which increases
with $\alpha_s(M_Z^2)$ as already discussed above.  Note that our obtained
value for $\mu_{\rm NLO}$ would have been larger if we had used the 
(inappropriate) approximate formula (2) instead of the exact solution
of (1).  It should be furthermore noted that we have chosen, as previously 
\cite{ref3}, a vanishing strange sea at the input scale $\mu$ in order to 
comply with experimental indications \cite{ref26,ref27} of an SU(3)--broken 
sea.  This choice is also supported and slightly favored by our input fits and 
compares well \cite{ref28} with recent measurements \cite{ref27} of $s(x,Q^2\! 
> \!\mu^2)$.  Our $s(x,Q^2)$ is thus generated purely dynamically (radiatively) 
and therefore constitutes, for the time being, an absolute, i.e.\ 
parameter--free prediction.  If future experiments may require a finite strange 
sea input, then our present results for $s(x,Q^2)$ have to be interpreted as an 
absolute lower bound for the strange sea.  Furthermore, the charm contribution 
$F_2^c(x,Q^2)$ to $F_2$ is provided by the perturbatively stable \cite{ref29} 
fixed--order perturbation theory.  In LO it derives from the well known 
photon--gluon fusion process \cite{ref3} $\gamma^*g\to c\bar{c}$.  For the NLO 
calculations we employ the $O(\alpha_s^2)$ coefficient functions of 
\cite{ref30} as conveniently parametrized in \cite{ref31}. In both cases we use 
$m_c=1.4$ GeV as given in (3) and choose the factorization and renormalization 
scales to equal $4m_c^2$.  The bottom contribution $F_2^b$ is marginal, 
reaching at most 1 to 2\%.

Our resulting small--$x$ predictions for $F_2^p$ are shown in Fig.\ 3 and for 
$xg\,\, (x,Q^2=20$ GeV$^2$) by the solid curve in Fig.\ 2.  These results are 
not too different from our NLO GRV(94) expectations \cite{ref3} when comparing 
them with the NLO(94) results in Figs.\ 1 \mbox{and 2}.  Moreover, the NLO 
results in Fig.\ 3 are favored over the LO ones in the small $Q^2$ region, 
$Q^2$ \raisebox{-0.1cm}{$\stackrel{<}{\ssim}$} 3 GeV$^2$ (which might indicate
that future NNLO contributions could even improve our NLO results in the small 
$Q^2$ region around 1 GeV$^2$).  Furthermore, their dependence on choosing 
different factorization scales \cite{ref14,ref32}, instead of $\mu_F^2=Q^2$, is 
obviously weaker than in LO.  Nevertheless, the present LO/NLO stability is
even better than for the GRV(94) $F_2$--predictions.  The LO and NLO input 
densities in eqs.~(5) and (6) are shown in Fig.~4 and compared with the NLO 
ones of GRV(94) \cite{ref3} as well as with their evolutions to $Q^2=5$ 
GeV$^2$.  The appropriate valence--like NLO input $F_2^p(x, \mu_{\rm NLO}^2)$, 
which eventually vanishes as $x\to 0$, is also shown in the lower right corner 
of Fig.\ 3.  This illustrates very clearly that the predictions at $x < 
10^{-2}$ and $Q^2> \mu_{\rm NLO}^2$ are of a purely dynamical origin, in 
particular the increase of $F_2$ with $x$ as $x\to 0$ is due to the 
non--vanishing input at \mbox{$x$ \raisebox{-0.1cm}{$\stackrel{>}{\ssim}$} 
$10^{-2}$}.  Nevertheless, as evident from Fig.\ 3, our predictions for $Q^2$ 
\raisebox{-0.1cm}{$\stackrel{<}{\ssim}$} 1 GeV$^2$ fall below the data in the 
(very) small--$x$ region.  This is not unexpected for leading twist--2 results, 
since nonperturbative (higher twist) contributions to $F_2(x,Q^2)$ have 
eventually to become dominant for decreasing values of $Q^2$.  

It is also interesting to note that the total momentum fractions carried by the 
NLO input distributions at $Q^2=\mu^2$ in Fig.\ 4 amount to 56, 30 and 14\% for 
valence, gluon and sea densities, respectively, which are very similar in LO 
and similar to our GRV(94) results \cite{ref3}. 

Further typical small--$x$ predictions for $xg(x,Q^2)$ and $x\bar{u} (x,Q^2)$ 
are shown in Fig.\ 5 together with their respective inputs at $Q^2=\mu^2$ which 
become, particularly for the gluon input, vanishingly small at $x<10^{-2}$.  
This illustrates again the purely dynamical origin of the small--$x$ structure 
of gluon and sea quark densities at $Q^2>\mu^2$.  Also noteworthy is the 
stability of $\bar{u}(x,Q^2)$ at $Q^2\gg \mu^2$, i.e.\ not only the 
perturbative LO/NLO one but also that with respect to our GRV(94) results.  
This stability is almost as good as the one required for a physical quantity 
like $F_2(x,Q^2)$.  The situation is, as usual \cite{ref3}, different for 
$g(x,Q^2)$ which is, however, not as relevant since the gluon density is not 
directly measurable.  In fact, despite the sizeable difference of the LO and
NLO gluon distributions in Fig.\ 5 in the small--$x$ region, the directly
measurable gluon--dominated heavy quark contribution $F_2^c(x,Q^2)$ shows a 
remarkable perturbative stability even for very large values of the
 (factorization) scale, such as $\mu_F^2\sim Q^2$, as will be shown in the next 
Section.  Furthermore, $xg$ and $x\bar{u}$ at $Q^2\gg \mu^2$ increase almost 
linearly for $10^{-5}$ \raisebox{-0.1cm}{$\stackrel{<}{\ssim}$} $x$ \raisebox
{-0.1cm}{$\stackrel{<}{\ssim}$} $10^{-3}$ and $10^{-5}$ \raisebox{-0.1cm}
{$\stackrel{<}{\ssim}$} $x$ \raisebox{-0.1cm}{$\stackrel{<}{\ssim}$} $10^{-2}$,
respectively, on the double--logarithmic plots in Fig.\ 5.  They can thus 
effectively be represented by $xf(x,Q^2) \sim x^{-\lambda_f(x,Q^2)}$ with 
effective slopes
\begin{eqnarray}
\lambda_g(x,5\,{\rm GeV}^2)\simeq 0.24\,\, (0.34), &\quad\quad &
  \lambda_g(x,20\,{\rm GeV}^2)\simeq 0.30\,\, (0.39)\nonumber\\
\lambda_{\bar{u}}(x,5\, {\rm GeV}^2)\simeq 0.21\,\, (0.19), & \quad\quad &
  \lambda_{\bar{u}}(x,20\, {\rm GeV}^2)\simeq 0.26\,\, (0.27)\, ,
\end{eqnarray}
in NLO(LO), valid in the above mentioned $x$--intervals.  These steep 
$(\lambda_f>0)$ dynamical small--$x$ predictions are somewhat smaller
than the ones of GRV(94) \cite{ref3}.

Fig.\ 6 is an alternative way to present our dynamical (steep) small--$x$
predictions using the slope of $F_2^p(x,Q^2)$.  This directly measurable gluon 
dominated quantity, $dF_2^p/d\, \ln Q^2$, exhibits again a good perturbative 
stability for $Q^2$ \raisebox{-0.1cm}{$\stackrel{>} {\ssim}$} 1 GeV$^2$. The 
similarity of our new (modified) results with our previous NLO GRV(94) ones 
\cite{ref3} is interesting as well.  We refrain here from plotting any 
experimental results which are usually $Q^2$--averaged \cite{ref33,ref34} and 
thus may easily give rise to erroneous and misleading interpretations and 
conclusions when compared with theoretical results for $dF_2^p(x,Q^2)/d\, 
\ln Q^2$ which depend strongly on the specific choices of $x$ and $Q^2$. 
 Nevertheless, our slopes in Fig.\ 6 are consistent with present HERA 
measurements \cite{ref33,ref34} for not too small values of $Q^2$.

\section{The role of further deep inelastic and hard \mbox{scattering}
 data and very small--\boldmath{$x$} predictions}
\vspace*{-2mm}

As mentioned in the Introduction and the previous Section, fixed--target DIS 
measurements and data on $F_2^c(x,Q^2)$, $pp(n) \to \mu^+\mu^-X$ and the 
$p\bar{p}\to W^{\pm}X$ charge asymmetry are relevant for fixing and testing 
$g(x,Q^2),\, \,\bar{d}-\bar{u}$ and $d/u$.  In Figs.\ 7 and 8 we compare our 
fitted LO and NLO results with the fixed--target $F_2^p$ data\footnote{The 
normalizations of the $F_2^p$ data sets are allowed to
float within their experimental uncertainties; the resulting normalization
factors are 1.00 (SLAC), 0.98 (BCDMS), 1.01 (NMC, E665).  The BCDMS data have 
been taken as analyzed in \cite{ref19}, i.e., with a shift of the central 
values due to the main systematic error and correspondingly reduced full 
errors.  This is still the consistent treatment, also in conjunction with the 
new NMC data \cite{ref19}.  Furthermore we take over the target--mass 
corrections of \cite{ref19} for the SLAC and BCDMS $F_2^p$ data.}
[17--21] where we show only the kinematic region which has
been used to determine our valence--like gluon and sea $(\bar{u}+\bar{d})$ and 
the valence input distributions at $Q^2=\mu_{\rm LO,NLO}^2$ as given in 
eqs.~(5) and (6).  The quality of our NLO fits (which are slightly favored over 
their LO counterparts) in Fig.\ 7 clearly demonstrates that our chosen value of 
$\alpha_s(M_Z^2)=0.114$ in Section~2 is fully consistent with all present 
fixed--target high precision DIS (non--singlet) data in the large--$x$ region. 
In our present analysis we employ the $F_2^n/F_2^p$ data \cite{ref22,ref23} as 
extracted without any (still uncertain) corrections for a possible EMC effect 
and nuclear shadowing in deuterium.  Together with the $d_v/u_v$ constraints 
from $\nu p$ and $\nu d$ data \cite{ref24}, this procedure leads to a 
`traditional' large--$x$ behavior, $d_v/u_v\to 0$ as $x\to 1$.  Note that we 
investigated more complicated parametrizations for $d_v(x,\mu^2)$ than used in 
(5) and (6), but found them unnecessary even at the present level of accuracy. 

In Fig.\ 9 we compare our new (modified) LO and NLO parton distributions with 
the relevant data on DIS charm production \cite{ref36,ref37,ref38,ref39}. From 
HERA only the most recent (preliminary) 1997 ZEUS results \cite{ref39} are 
displayed, since they supersede the previous (published) measurements 
\cite{ref37,ref38} by their greatly improved accuracy.
As discussed in the previous Section, the gluon $g(x,\mu_F^2)$ dominated 
$F_2^c(x,Q^2)$ is calculated using the fully predictive {\it fixed\/} order 
(LO/NLO) perturbation theory \cite{ref3,ref30} which also underlies the actual 
analysis of the partial data \cite{ref37,ref38} utilized for extracting the 
total $F_2^c(x,Q^2)$.  The factorization scale $\mu_F$ (being as usual assumed 
equal to the renormalization scale) should be preferably chosen to be 
$\mu_F^2=4\, m_c^2$ \cite{ref29}.  The resulting predictions in Fig.\ 9 are in 
perfect agreement with all available data (including the original fixed--target 
EMC data \cite{ref36}) and are furthermore perturbatively stable.  Even 
choosing a very large scale like $\mu_F^2=4(Q^2+4\, m_c^2)$, the NLO results 
remain essentially unchanged at small--$x$ \cite{ref40} as shown by the `high 
scale' dotted curves in Fig.\ 9.  This latter stability renders attempts to 
resum supposedly large logarithms ($\ln Q^2/m_c^2$) in heavy quark production
cross sections superfluous \cite{ref11,ref12,ref41}.  It should be noted that 
the charm (and bottom) production data strongly constrain the gluon
distribution and will eventually be used to determine $g(x,\mu_F^2)$ directly 
from experiment \cite{ref8}.

The asymmetry measurements of Drell--Yan dilepton production in $pp$ and $pd$ 
collisions \cite{ref9,ref10} which have been instrumental in fixing 
$\bar{d}-\bar{u}$ (or $\bar{d}/\bar{u}$), in particular the very recent 
Fermilab--E866 data \cite{ref10} for the extented $x$ range 0.03 \raisebox
{-0.1cm}{$\stackrel{<}{\ssim}$} $x$ \raisebox{-0.1cm} {$\stackrel{<}{\ssim}$} 
0.35, are compared with our updated LO and NLO results in Fig.\ 10.  For 
comparison the consequences of our NLO GRV(94) $\bar{d}/\bar{u}$ ratio are 
shown as well which has been originally constrained just by the CERN--NA51 
measurement \cite{ref9} at $x=0.18$.  Our present new results for $\bar{d}/
\bar{u}$ are also consistent with (although slightly lower than) the recent 
preliminary semi--inclusive HERMES measurements \cite{ref42} at $0.05 \leq x
\leq 0.2$.  The sensitivity of the Drell--Yan asymmetry on $\bar{d}-\bar{u}$ 
can be most easily seen from the LO expression
\begin{equation}
A_{DY} \equiv \frac{\sigma^{pp}-\sigma^{pn}}{\sigma^{pp}+\sigma^{pn}} =
  \frac{2\sigma^{pp}}{\sigma^{pd}} \, -\, 1 = 
  \frac{(u-d)\,(\bar{u}-\bar{d})\,+\,\frac{3}{5}\,(u\bar{u}-d\bar{d})}
  {(u+d)\,(\bar{u}+\bar{d})\,+\,\frac{3}{5}\,(u\bar{u}-d\bar{d})\,+
   \frac{4}{5}\,s\bar{s}}
\end{equation}
\\[-4mm]
due to $\sigma^{pN}\propto\sum_{u,d,s} e_q^2\left[ q(x_1)\bar{q}(x_2) +
q(x_2)\bar{q}(x_1) \right]$.
The relevant NLO differential Drell-Yan cross section $\sigma^{pN}\equiv
d^2\sigma^{pN}/dM_{\mu^+\mu^-}dx_F$ can be found in the Appendix of 
\cite{ref43}, except for eq.~(8a) which has to be modified \cite{ref44,ref45} 
in order to conform with the usual $\overline{\rm MS}$ convention for the 
number of gluon polarization states $2(1-\varepsilon)$ in $4-2\varepsilon$ 
dimensions.

Having fixed $\bar{d}-\bar{u}$ and $\bar{d}+\bar{u}$ as well as the 
valence densities $u_v$ and $d_v$ (cf.\ Figs.\ 7 and~8), our strongly
constrained $u$ and $d$ distributions are now confronted in Fig.\ 11
with the $W^{\pm}\to\ell^{\pm}\nu$ charge asymmetry measurements at the
Fermilab $p\bar{p}$ collider \cite{ref46}.  The $W^{\pm}$~rapidity
asymmetry
\begin{equation}
A(y_{\ell}) = \frac{ d\sigma(\ell^+)/dy_{\ell} - d\sigma(\ell^-)/dy_{\ell}}
  {d\sigma(\ell^+)/dy_{\ell} + d\sigma(\ell^-)/dy_{\ell}}
\end{equation}
\\[-4mm]
of the charged leptons from the $W^{\pm}\to\ell^{\pm}\nu$ decays with
the lepton rapidity $y_{\ell}$ is a sensitive probe of the difference
between $u$ and $d$ quark distributions at $Q^2=M_W^2$.  Our LO and NLO
predictions are in perfect agreement with present data.  A good agreement 
is also obtained \cite{ref47} with our GRV(94) densities \cite{ref3}, as 
shown in Fig.\ 11, and even with our previous $(\bar{u}=\bar{d})$ dynamical 
LO and NLO distributions \cite{ref1,ref2}.

Further constraints on the gluon distribution, besides those from DIS and
Drell--Yan data \cite{ref12}, could be obtained via $pp\to\gamma X, \,\, 
pp\to{\rm jet}+X$, etc., but the predicted cross sections are quite sensitive 
to assumptions concerning the \mbox{magnitude} of the intrinsic transverse 
momentum $k_T$ of the partons which is not well \mbox{understood} at present. 
 This holds true in particular for the prompt photon \mbox{production} data 
where, despite the large scale ($\mu_{R,F}$) uncertainty \cite{ref48,ref49}, 
an additional sizeable $k_T$--smearing seems to be required 
\cite{ref49,ref50,ref11} in particular in view of the recent Fermilab (E706) 
$pBe\to\gamma X$ measurements \cite{ref51}.  A further interesting source 
of information, sensitive to $g(x,Q^2)$, will be provided by the longitudinal 
structure function $F_L(x,Q^2)\equiv F_2-2xF_1$  as illustrated for example
in \cite{ref3}.  The available data on $F_L$ are unfortunately still of rather 
limited accuracy \cite{ref8,ref52}.  Finally it should be mentioned that the 
gluon density is also tested in a NLO determination of the strange sea quark 
density $s(x,Q^2)$ from neutrino induced charm production data, i.e.\ 
opposite--sign dimuon events originating from $\nu N\to\mu^-cX$ with 
$c\to\mu^+\bar{\nu}_{\mu}s$.  Since \mbox{dimuon} events give, among other 
things, direct access to $s(x,Q^2)$ via $W^+s\to c$, \mbox{$W^+s\to cg$} and 
$W^+g\to c\bar{s}$, etc., they also probe our purely dynamically generated 
$s(x,Q^2)$, i.e.\ the assumed vanishing strange sea input in eqs.~(5) and (6), 
throughout the entire $x$--region. Since our slightly modified input densities 
in (5) and (6) give rise to $s(x,Q^2)$ and $g(x,Q^2)$ which are similar to the 
GRV(94) ones, they result in similarly agreeable predictions for 
$\stackrel{{\scriptscriptstyle (-)}}{\nu}\!\!N\to\mu^-\mu^+X$ \cite{ref28}. 
It should, however, be kept in mind that a vanishing strange sea input is by no 
means a crucial ingredient of the dynamical approach.

As our parameter--free small--$x$ predictions for parton distributions at 
$x<10^{-2}$ are entirely of QCD--dynamical origin and depend, apart from 
intrinsic theoretical uncertainties, rather little on the detailed input 
parameters at $x$ \raisebox{-0.1cm}{$\stackrel{>}{\ssim}$} $10^{-2}$, it is 
interesting to study these predictions in kinematic regions not accessible by 
present DIS experiments and to compare them with our previous GRV(94) 
densities.  Of particular interest here is the comparison in Fig.\ 12 at 
$Q^2=10^4$ GeV$^2$ and extremely small $x$, i.e.\ $10^{-8}$ \raisebox{-0.1cm}
{$\stackrel{<}{\ssim}$} $x$ \raisebox{-0.1cm}{$\stackrel{<} {\ssim}$} 
$10^{-5}$, relevant to questions concerning neutrino astronomy \cite{ref53}.  
These results at \mbox{$x$ \raisebox{-0.1cm}{$\stackrel{<}{\ssim}$} $10^{-5}$} 
indicate that, for example, ultra--high--energy neutrino nucleon cross 
sections, which are sensitive to parton densities at $x$ values as small as 
$10^{-8}$, can be rather reliably calculated to \mbox{within} about 20\%.  This 
follows not only from the perturbative stability at extremely small values of 
$x$, where moreover our improved predictions are comparable to the ones based 
on the NLO GRV(94) densities, but also from the fact that the predictions at 
$Q^2\simeq M_W^2$ are rather independent of the specific choice for the 
renormalization scale $\mu_R$ appearing in $\alpha_s(\mu_R^2)$ and for the 
factorization scale $\mu_F$ appearing in the parton densities $f(x,\mu_F^2)$. 
 We have checked this by taking $\mu_R=\mu_F$ with \cite{ref14,ref32} $Q/2\leq 
\mu_F\leq 2Q$ which requires, of course, also corresponding 
modifications\footnote{The implementation of this
modification amounts to replacing everywhere in \cite{ref3,ref54} the NLO 
$\alpha_s(Q^2)$ and the $f(x,Q^2)$ by $\alpha_s(\mu_F^2)$ and $f(x,\mu_F^2)$, 
respectively, while the common $\overline{\rm MS}$ Wilson coefficients in 
\cite{ref3} have to be replaced by $C_i\to C_i+P_{qi}^{(0)}\, \ln Q^2/\mu_F^2$ 
for $i=q,\, g$.} in our input scale $\mu$ and $f(x,\mu^2)$ in eq.~(6). 
 
\vspace*{-2mm}
\section{Discussion and Summary}
\vspace*{-2mm}

As demonstrated above, the radiative (dynamical) GRV(94) parton distributions 
\cite{ref3} dis-\- agree with recent precision HERA data only within the 
margins resulting from the 10\% uncertainty in their input scale $\mu$ 
\cite{ref2,ref3}.  Taking into account also the new large--$x$ parton 
constraints and $\alpha_s$--results, we have generated new sets of LO and NLO 
dynamical parton densities corresponding to $\alpha_s(M_z)= 0.114$ for the 
purpose of future precision analyses.  Nevertheless one can in practice still 
utilize the former GRV(94) distributions \cite{ref3}, in particular in view of 
the fact that in most applications of these parton densities to, say, 
high--$p_T$ jet, photon or heavy quark production, the usually considered 
(see, for example, \cite{ref14,ref32} for a recent review and comparative 
discussion) freedom in the choice of the factorization and/or renormalization 
scale, e.g.\ $p_T/2$ \raisebox{-0.1cm}{$\stackrel{<}{\ssim}$} $\mu_{F,R}$ 
\raisebox{-0.1cm}{$\stackrel{<}{\ssim}$} $2\, p_T$, overshadows the present 
modifications of our previous distributions \cite{ref3}.

It is also necessary to mention that the task of searching the ultimately
correct parton distributions is not only affected by the above mentioned
higher order uncertainties, but also by the discrepancies between the 
data sets used, e.g.\ between the NMC data \cite{ref20} on the one hand
and the CCFR \cite{ref55} and HERA \cite{ref6,ref7,ref8} data on the 
other, which are not well understood at present \cite{ref33,ref56}.
Furthermore, recent attempts to calculate the (non--perturbative) input
parton densities from first principles using the chiral soliton approach
yielded, besides the valence densities, also a {\it valence--like\/} sea
density in the small--$x$ region at $Q_0^2=0.3-0.4$ GeV$^2$ -- a scale
set by the inverse average instanton size \cite{ref57}.  It remains to
be seen, however, whether a sizeable valence--like gluon density at the
same `dynamical' input scale is also within the realm of this approach.
It is also interesting to remark that a valence--like gluon input density,
with a momentum fraction compatible with our results \cite{ref1,ref2,ref3}, 
has been obtained from considerations of intrinsic nucleon Fock--states 
\cite{ref58}.

Finally it should be emphasized that the stable parameter--free dynamical
predictions for parton distributions in the extremely small--$x$ region,
$10^{-8}$ \raisebox{-0.1cm}{$\stackrel{<}{\ssim}$} $x$ \raisebox{-0.1cm}
{$\stackrel{<}{\ssim}$} $10^{-5}$, allow for rather reliable estimates of 
ultra--high--energy neutrino nucleon cross sections relevant to questions 
concerning neutrino astronomy \cite{ref53}.

A FORTRAN package containing our new LO and NLO($\overline{\rm MS}$) parton 
densities as well as $F_2^{c,b}(x,Q^2)$, calculated in fixed--order
perturbation theory, can be obtained by electronic mail on request.
Instead of using the appropriate massive quark subprocesses for calculating
heavy quark production rates in fixed--order perturbation theory, rough
estimates (valid to within a factor of 2, say) of `heavy' quark effects can 
be easier obtained with the help of the massless `heavy' quark distributions 
$c(x,Q^2)$ and $b(x,Q^2)$ given in \cite{ref1}. For further convenience we 
also provide the NLO(DIS) distributions which are related to the 
NLO($\overline{\rm MS}$) ones according to eq.~(21) of ref.~\cite{ref3}.
 
\vspace*{-2mm}
\section*{Acknowledgements} 
\vspace*{-2mm}

We thank S. Kretzer for several helpful numerical analyses, in particular
for his cooperation in calculating the $W^{\pm}$ charge asymmetries in Fig.~11, 
as well as A. Milsztajn, \mbox{L. Bauerdick}, A. Bruell, E. Rondio,
V. Shekelyan, M. Vincter, and R.G. Roberts for useful discussions and 
informations concerning recent data.  This work has been supported in part by 
the "Bundesministerium f\"ur Bildung, Wissenschaft, Forschung und Technologie", 
Bonn.

\newpage
\setlength{\baselineskip}{0.64cm}

\newpage

\setlength{\baselineskip}{0.635cm}
\section*{$\!\!\!\! $Figure Captions}
\vspace*{-2mm}
\begin{itemize}
\item[\bf{Fig.\ 1}]  Comparison of our NLO GRV(94) small--$x$ predictions
	\cite{ref3} for the proton structure function $F_2^p$ with recent
  	precision measurements at HERA \cite{ref6,ref7}.  For illustration 
	we also include the most recent preliminary data \cite{ref8} (open
	symbols).  The typical uncertainties of the GRV(94) predictions are
	illustrated by the dotted curves which are obtained by increasing
	$\mu_{\rm NLO}$ by 10\%, keeping everything else (valence--like
	input densities, $\alpha_s$, etc.) unchanged.  The results of our
	present new analysis, where all these input quantities are 
	consistently modified, lie in between the curves shown at small--$x$.
\item[\bf{Fig.\ 2}]  As in Fig.\ 1, but for the NLO gluon density at 
	$Q^2=20$ GeV$^2$.  The result of our present new analysis is shown
	as well (solid curve).  The shaded bands represent the preliminary
	experimental small--$x$ constraints as extracted from 
	$F_2$--measurements at HERA and the four data points are derived
	from deep--inelastic inclusive charm--production as analyzed by
	H1 \cite{ref8}.
\item[\bf{Fig.\ 3}]  Comparison of our new LO and NLO small--$x$ results
	for $F_2(x,Q^2)$, arising from the inputs (5) and (6), with HERA
	data for $Q^2$ \raisebox{-0.1cm}{$\stackrel{>}{\ssim}$} 1 GeV$^2$
	\cite{ref6,ref7,ref8}.  The valence--like NLO input, according to
	Fig.\ 4, is shown by the curve 	$(\mu_{\rm NLO}^2)$ at the lower
	right corner.  To ease the graphical representation we have plotted
	$F_2(x,Q^2)+i(Q^2) \times 0.5$, with $i$ indicated in the figure.
	The published data \cite{ref6,ref7} (closed symbols) are, different
	from Fig.\ 1, shown with small normalization changes as obtained in
	our fits:  H1(94)\raisebox{-0.1cm}{*}0.99, 
        ZEUS(94)\raisebox{-0.1cm}{*}1.01, shifted vertex data [ZEUS(94) and 
        H1(95)]\raisebox{-0.1cm}{*}0.97.
\item[\bf{Fig.\ 4}]  The valence--like LO and NLO input densities 
	$xf$ $(f=u_v,\, d_v,\, \bar{u},\, \bar{d},\, g)$ according to 
	eqs.~(5) and (6) at $Q^2=\mu_{\rm LO}^2=0.26$ GeV$^2$ and 
	$Q^2=\mu_{\rm NLO}^2=0.40$ GeV$^2$. \mbox{The strange} sea $s=\bar{s}$
	vanishes at the input scales $Q^2=\mu_{\rm LO,NLO}^2$.  The NLO
	GRV(94) input \cite{ref3} is also shown for comparison, as well
	as the evoluted results at $Q^2=5$ GeV$^2$.
\item[\bf{Fig.\ 5}]  The small--$x$ behavior of our radiatively generated
	gluon and sea--quark distributions in LO and NLO.  The valence--like
	inputs, according to eqs.~(5) and (6) as presented in Fig.\ 4, are
	shown by the lowest curves referring to $\mu^2$ for illustration.
	For comparison we also show the NLO GRV(94) predictions \cite{ref3}.
	The results are multiplied by the numbers indicated in brackets.   
\item[\bf{Fig.\ 6}]  The predicted slope $dF_2^p(x,Q^2)/d\, \ln Q^2$ in
	LO and NLO.  The difference between the dotted and solid curves is
	due to the NLO charm contribution $dF_2^c(x,Q^2)/d\, \ln Q^2$.
	For comparison we also show the NLO results of GRV(94) \cite{ref3}.
	The upper part of the figure refers to the values of $Q^2$ 	
	appropriate to the HERA measurements \cite{ref33,ref34,ref11}, 
	and the lower one to a representative fixed value of $Q^2$.
\item[\bf{Fig.\ 7}]  Comparison of our LO and NLO fits with fixed--target
	$F_2^p$ data \cite{ref17,ref18,ref19,ref20,ref21} in the $Q^2\geq 4$ 
        GeV$^2$, $W^2\geq 10$ GeV$^2$ region which were used for determining 
        the valence and valence--like gluon and sea input in (5) and (6). 
        The data sets are shown with their normalization factors as obtained 
        in the fit.  Some points with large errors, e.g.\ E665 at large--$x$, 
        are omitted.
\item[\bf{Fig.\ 8}]  Same as in Fig.\ 7 but for the measured $F_2^n/F_2^p$
	and $d_v/u_v$ ratios \cite{ref22,ref24}.  Only the NMC results for
	$F_2^n/F_2^p$ are shown, as they are much more accurate than the 
	corresponding BCDMS and E665 data \cite{ref23}.  The preliminary
	semi--inclusive HERMES data for $d_v/u_v$ are displayed as well
	\cite{ref35}.  For comparison our previous NLO GRV(94) results
	\cite{ref3} are also shown.
\item[\bf{Fig.\ 9}]  LO and NLO predictions for $F_2^c(x,Q^2)$ in 
	fixed--order perturbation theory ($\gamma^*$-gluon fusion, etc.)
	based on our new LO and NLO parton densities using 
	$\mu_F^2=4\, m_c^2$, $m_c=1.4$ GeV, compared with data from EMC
	\cite{ref36} and ZEUS \cite{ref39}.  The NLO results based on the
	NLO GRV(94) parton densities \cite{ref3} are close to the present
	LO(NLO) curves at large (small) values of $x$.  The NLO (high scale)
	curves refer to a significantly larger factorization scale,
	$\mu_F^2=4\, (Q^2+4\, m_c^2)$.
\item[\bf{Fig.\ 10}]  LO and NLO QCD results for $\sigma^{pd}/2\sigma^{pp}$
	for $A_{DY}$ in eq.~(8) compared with the NA51 \cite{ref9} and the
	most recent E866 \cite{ref10} Drell--Yan dimuon production data.
	The experimental (E866) normalization uncertainty of 0.01 has been
	used in the fit.  The NLO(94) curves correspond to the $\bar{d}/
	\bar{u}$ GRV(94) ratio \cite{ref3} which has been originally
	extracted just from the the NA51 data point at $x=0.18$ shown in
	the right figure.
\item[\bf{Fig.\ 11}]  LO and NLO predictions for the $W^{\pm}$ charge
	asymmetry $A(y_{\ell})$ in eq.~(9). Note that these data have {\it 
        not\/} been used for fixing our new input distributions.  The NLO(94)
	predictions, based on the GRV(94) densities \cite{ref3}, are shown
	for comparison.  The relevant LO/NLO expressions for $A(y_\ell)$
	can be found in \cite{ref47} (and references therein). The
	Fermilab--CDF data are taken from \cite{ref46}.
\item[\bf{Fig.\ 12}]  Predictions for $F_2^p(x,Q^2)$ for extremely small
	values of $x$.  The difference between the dotted and solid curves
	is due to the NLO heavy quark (charm, bottom) contributions, which
	derive from photon--gluon (quark) fusion processes.  The NLO(94)
	results correspond to the parton densities of \cite{ref3}.  The
	NLO results at these large values of $Q^2$ are insensitive to the
	specific choice of the factorization scale $\mu_F$.
\end{itemize}

\newpage
\setlength{\baselineskip}{0.65cm}
\setlength{\parskip}{0.0cm}
\addtolength{\topmargin}{-1.0cm}
\addtolength{\textheight}{1.0cm}

\vspace*{\fill}
\centerline{\epsfig{file=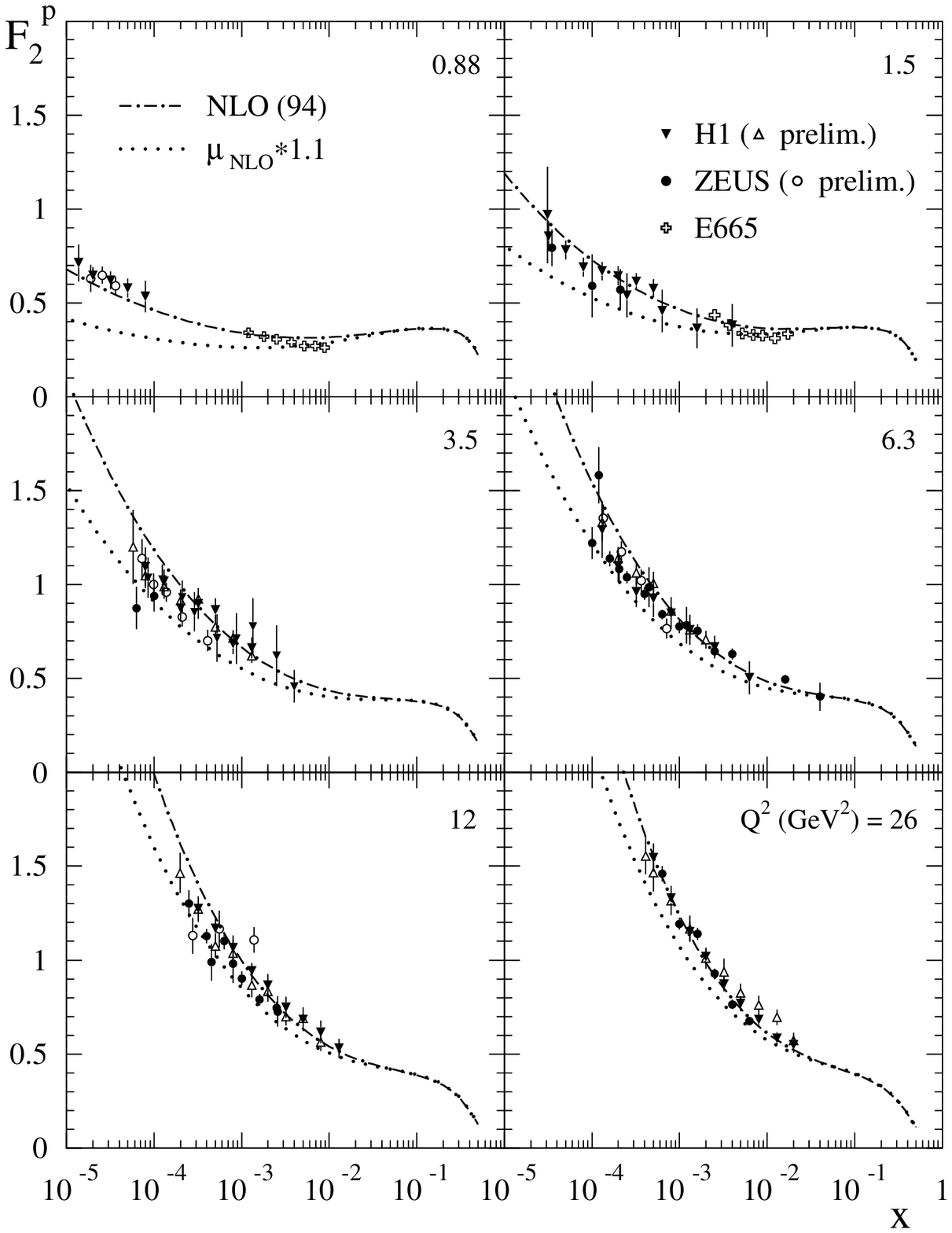,height=20cm}}
\vspace{3mm}
\centerline{\large\bf Fig.\ 1}
\vfill

\newpage
\vspace*{\fill}
\centerline{\epsfig{file=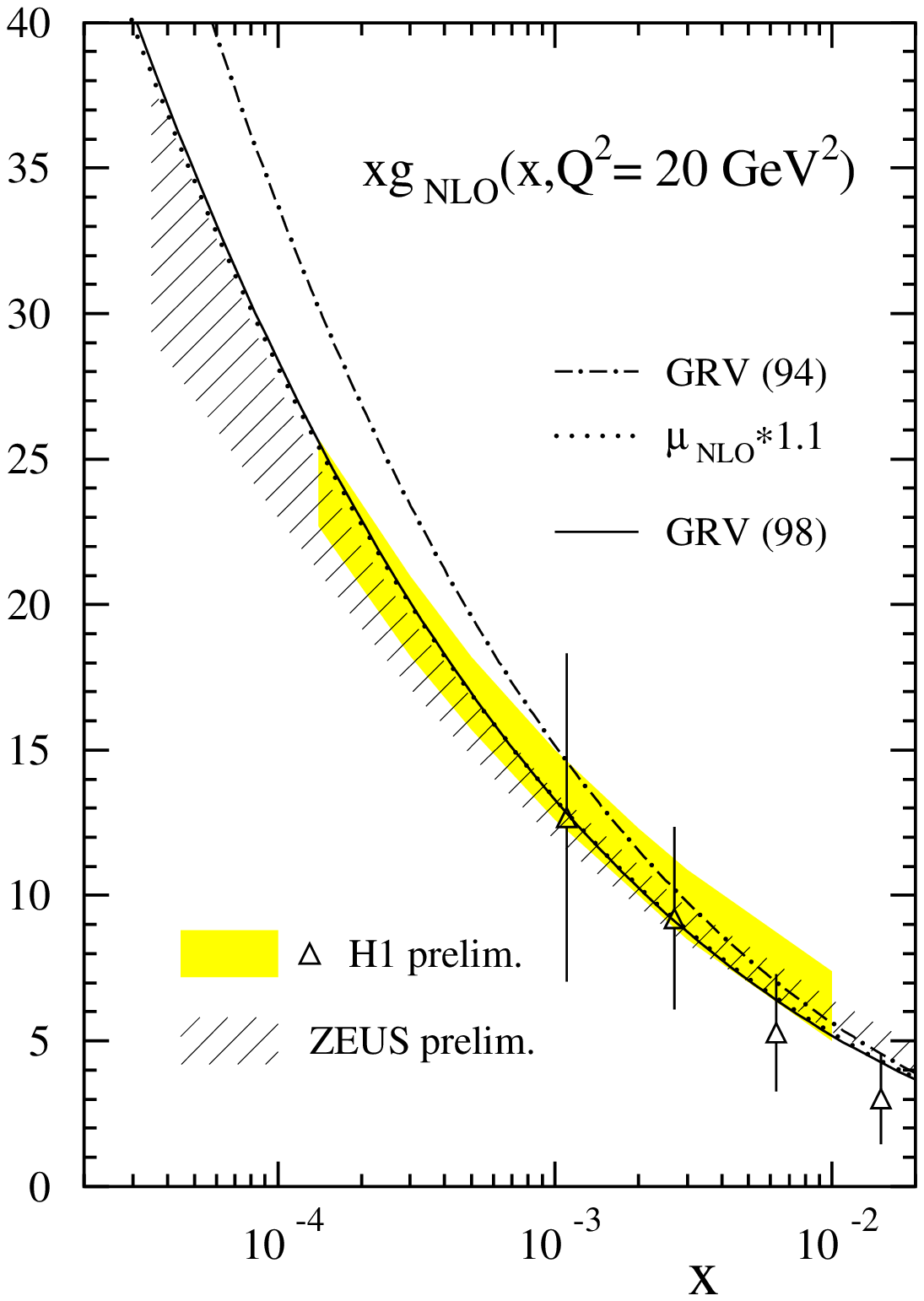,height=16cm}}
\vspace{3mm}
\centerline{\large\bf Fig.\ 2}
\vfill

\newpage
\centerline{\epsfig{file=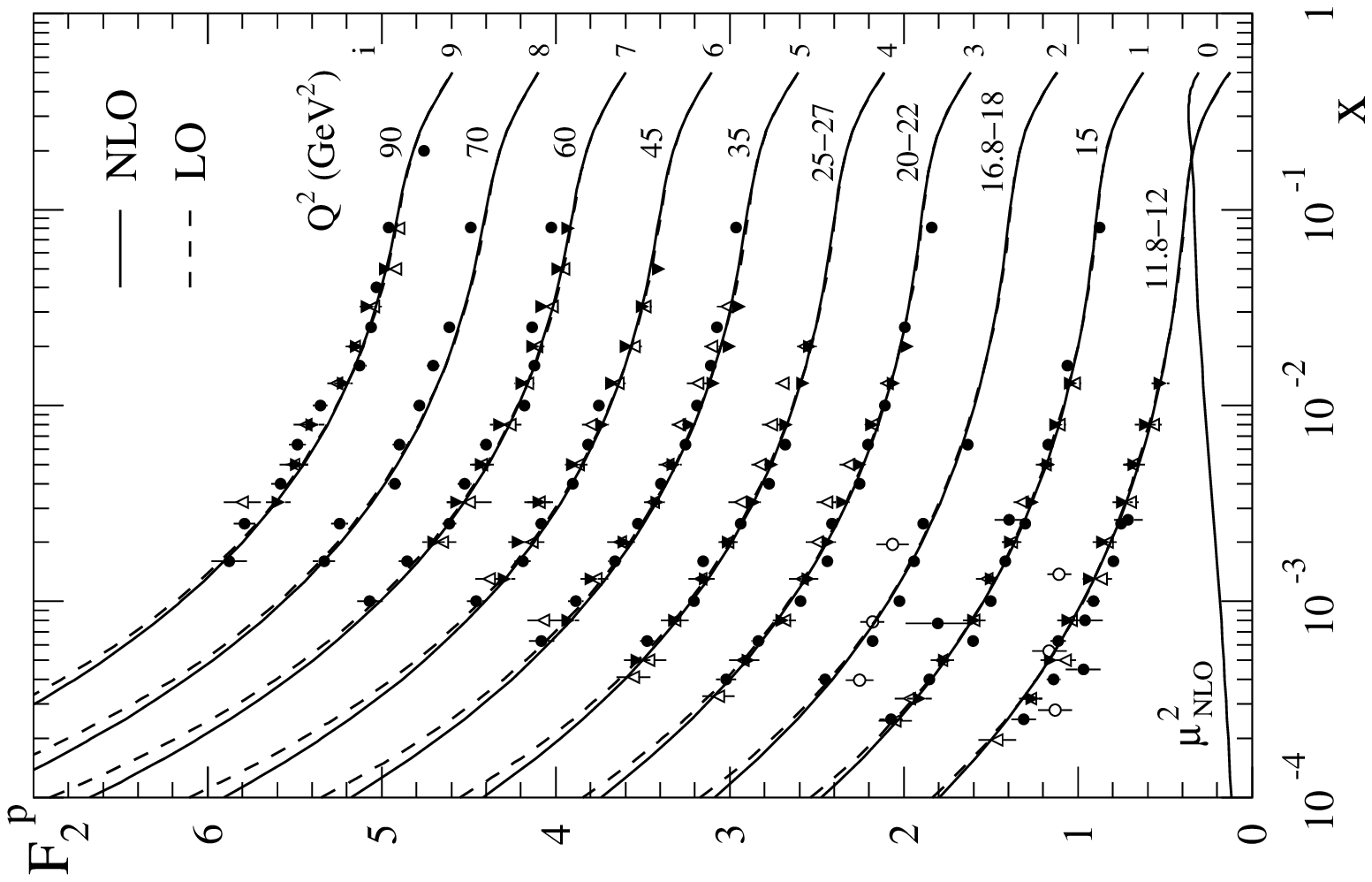,height=16cm,angle=90}}
 
\vspace*{-5mm}
\centerline{\epsfig{file=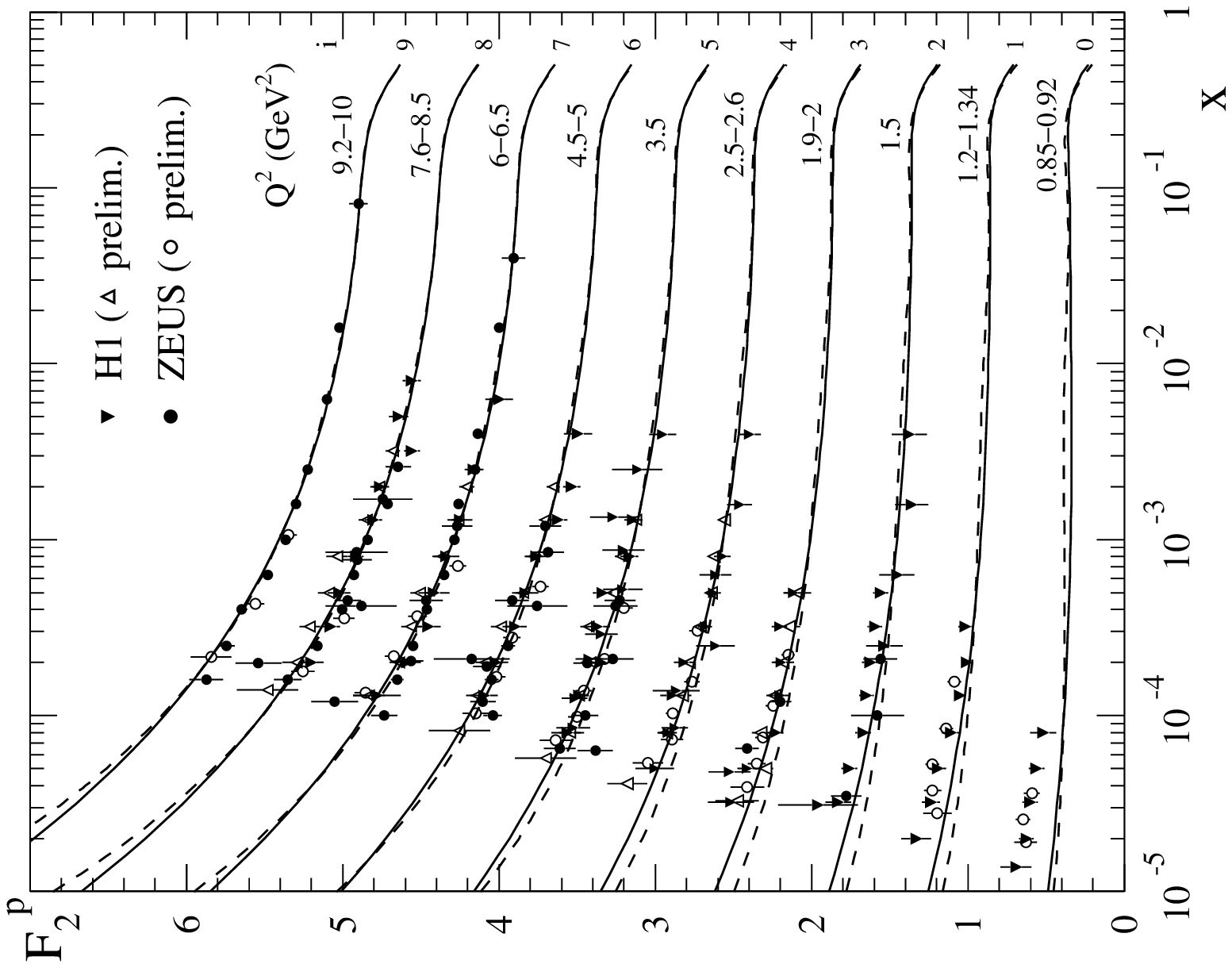,height=16cm,angle=90}}
\vspace{3mm}
\centerline{\large\bf Fig.\ 3}

\newpage
\vspace*{\fill}
\centerline{\hspace*{3mm}\epsfig{file=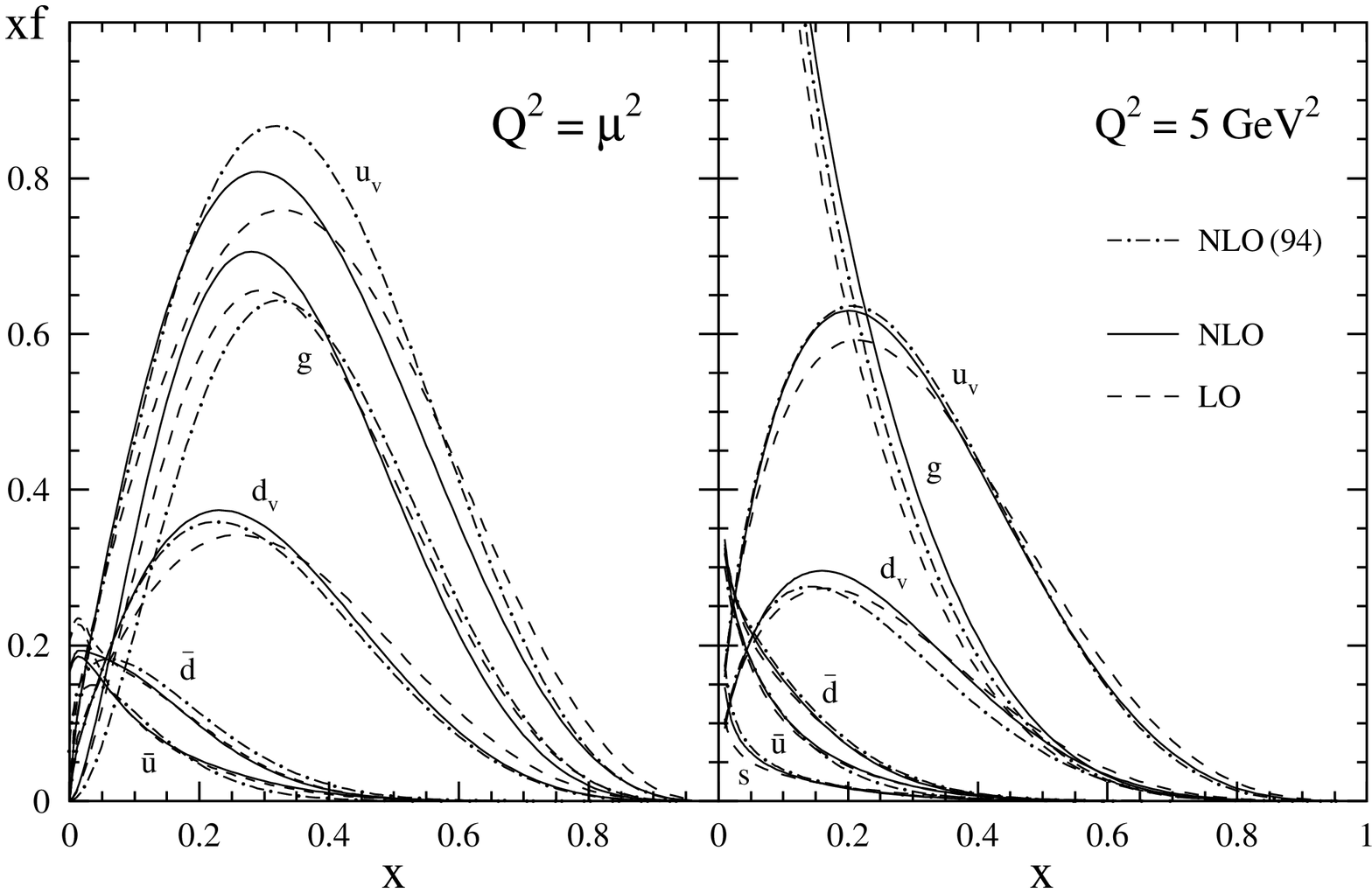,height=14cm,angle=90}}
\vspace{3mm}
\centerline{\large\bf Fig.\ 4}
\vfill

\newpage
\vspace*{\fill}
\centerline{\hspace*{3mm}\epsfig{file=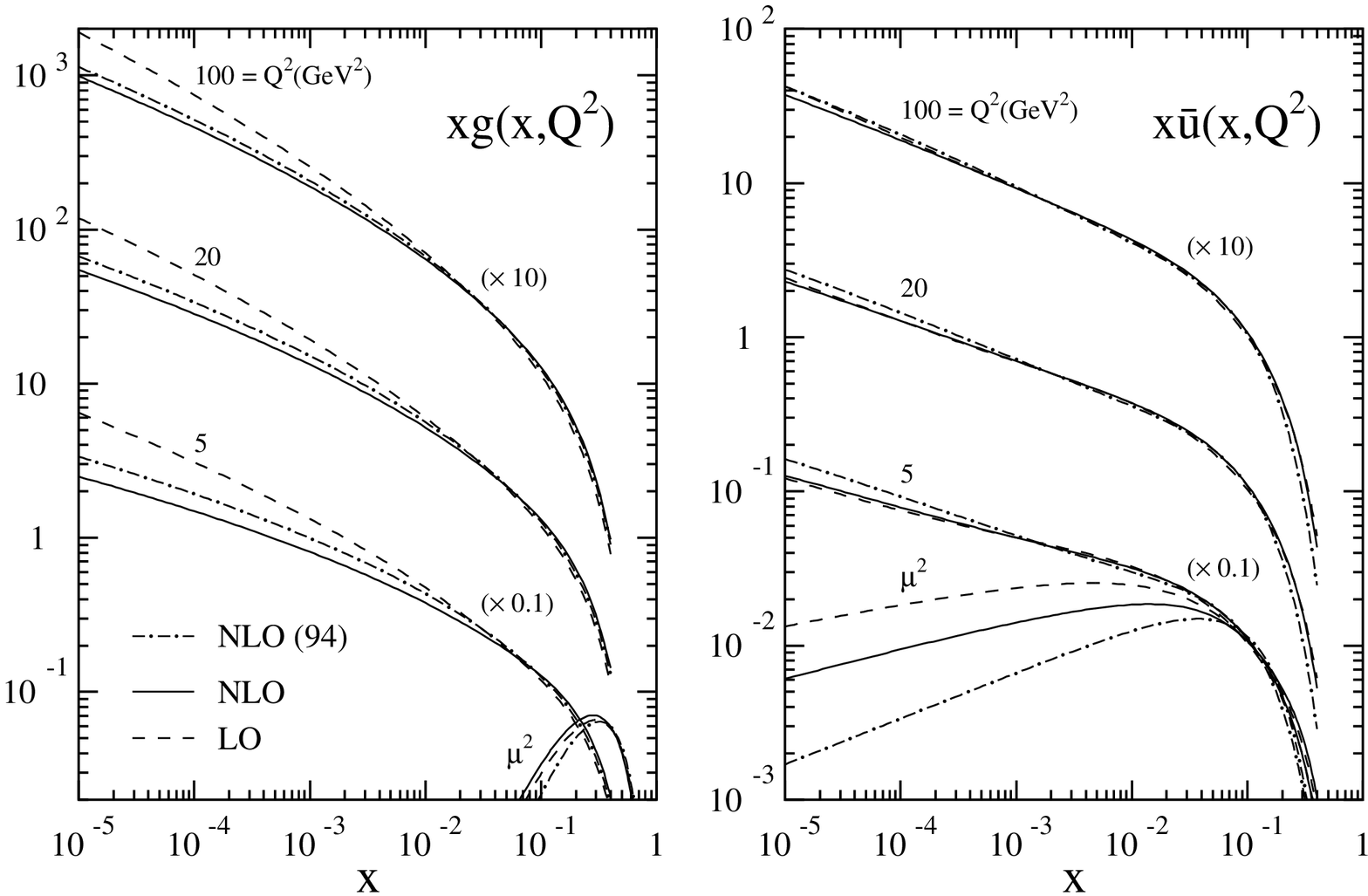,height=14cm,angle=90}}
\vspace{3mm}
\centerline{\large\bf Fig.\ 5}
\vfill

\newpage
\vspace*{\fill}
\centerline{\epsfig{file=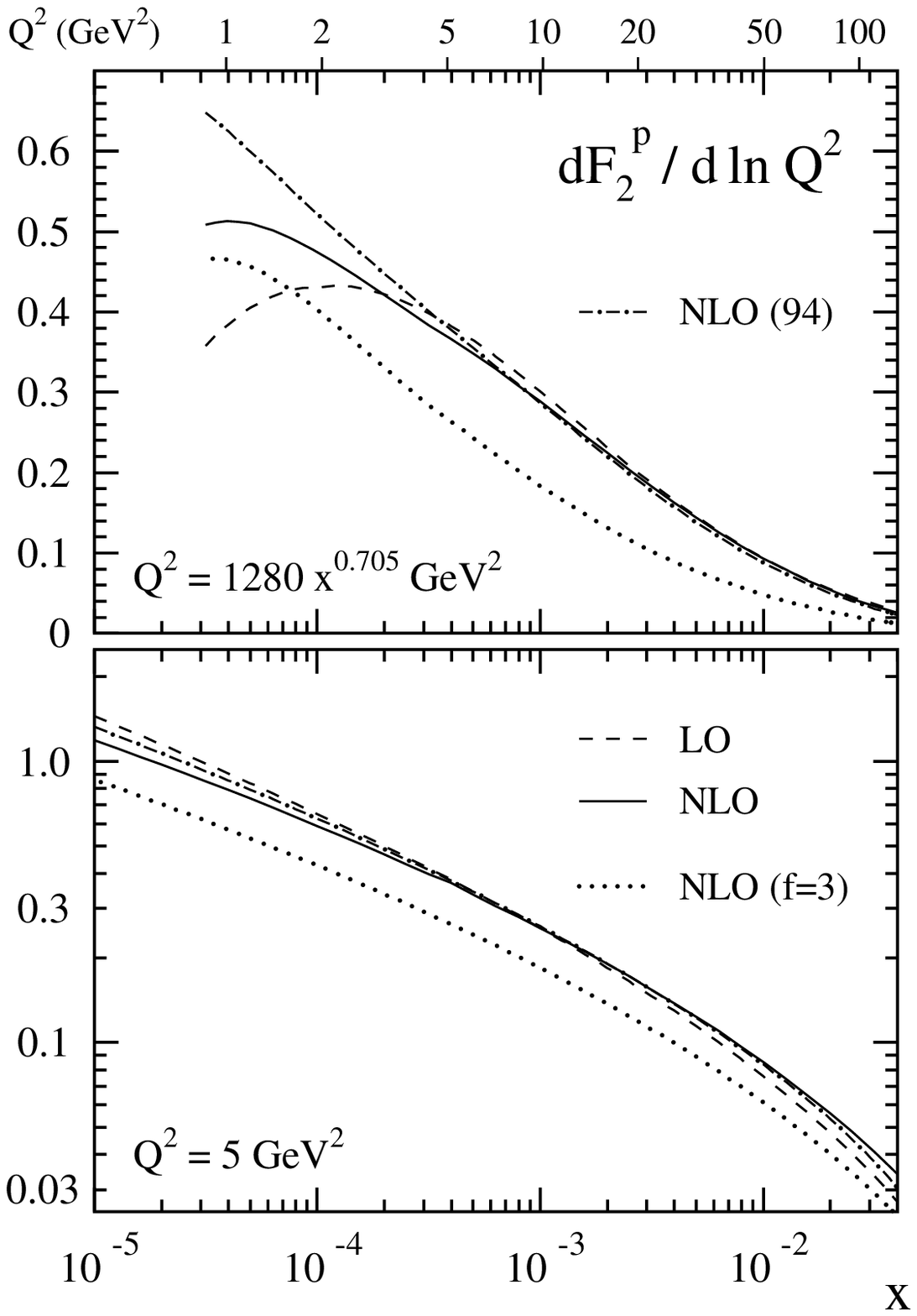,height=16.7cm}}
\vspace{3mm}
\centerline{\large\bf Fig.\ 6}
\vfill

\newpage
\vspace*{\fill}
\centerline{\hspace*{3mm}\epsfig{file=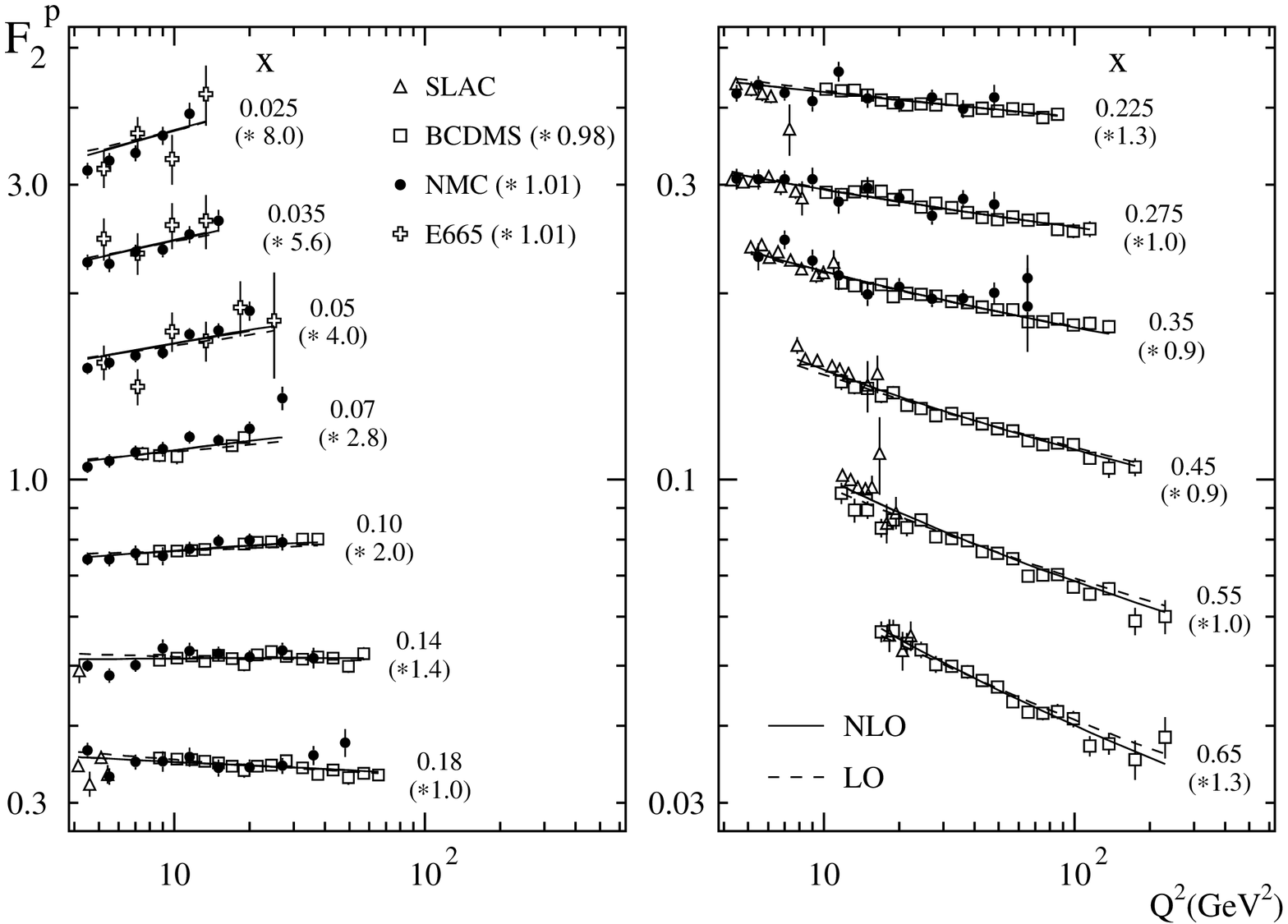,height=15cm,angle=90}}
\vspace{3mm}
\centerline{\large\bf Fig.\ 7}
\vfill

\newpage
\vspace*{\fill}
\centerline{\hspace*{3mm}\epsfig{file=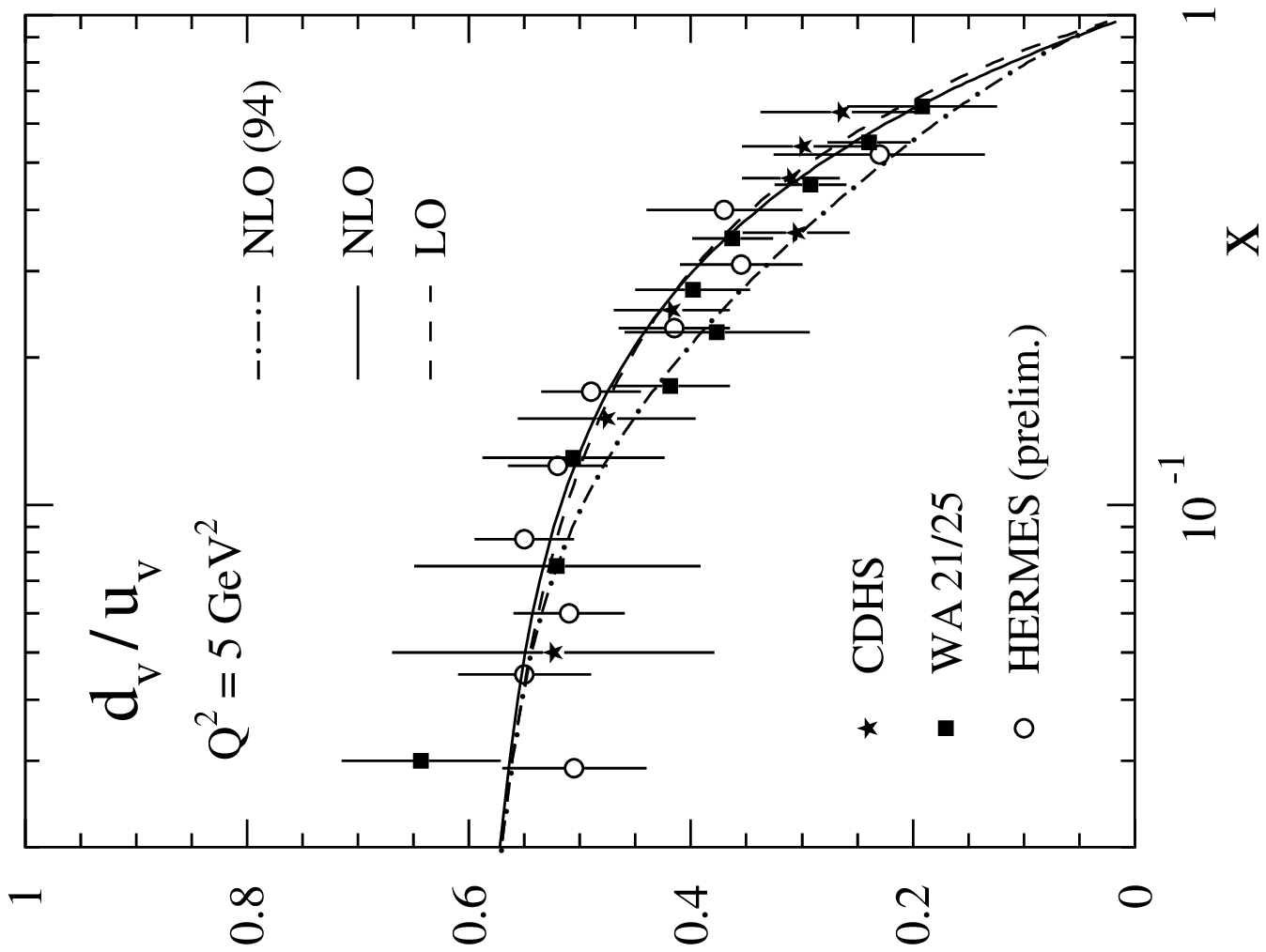,height=14cm,angle=90}}

\centerline{\hspace*{3mm}\epsfig{file=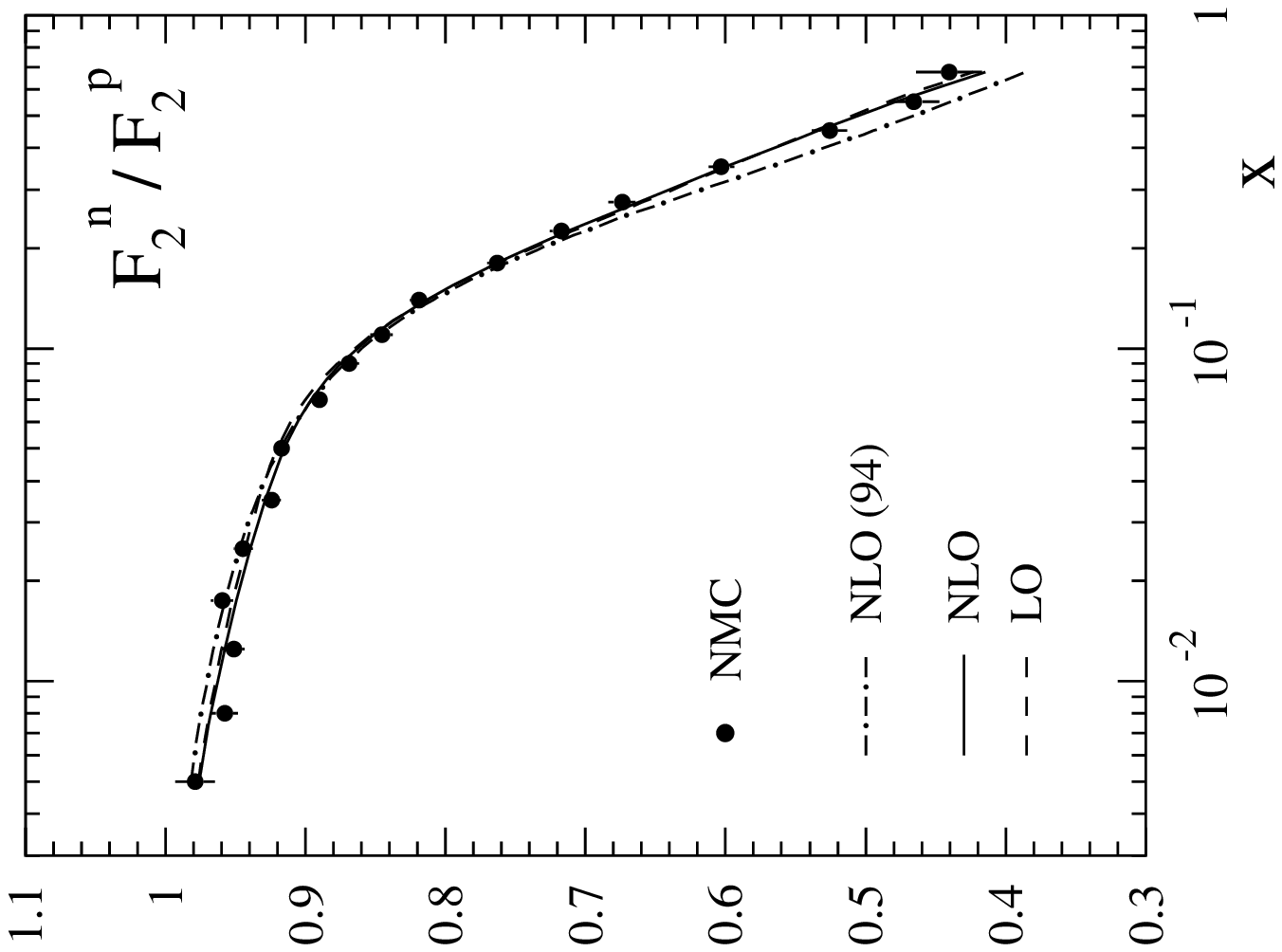,height=14cm,angle=90}}
\vspace{3mm}
\centerline{\large\bf Fig.\ 8}
\vfill

\newpage
\vspace*{\fill}
\centerline{\epsfig{file=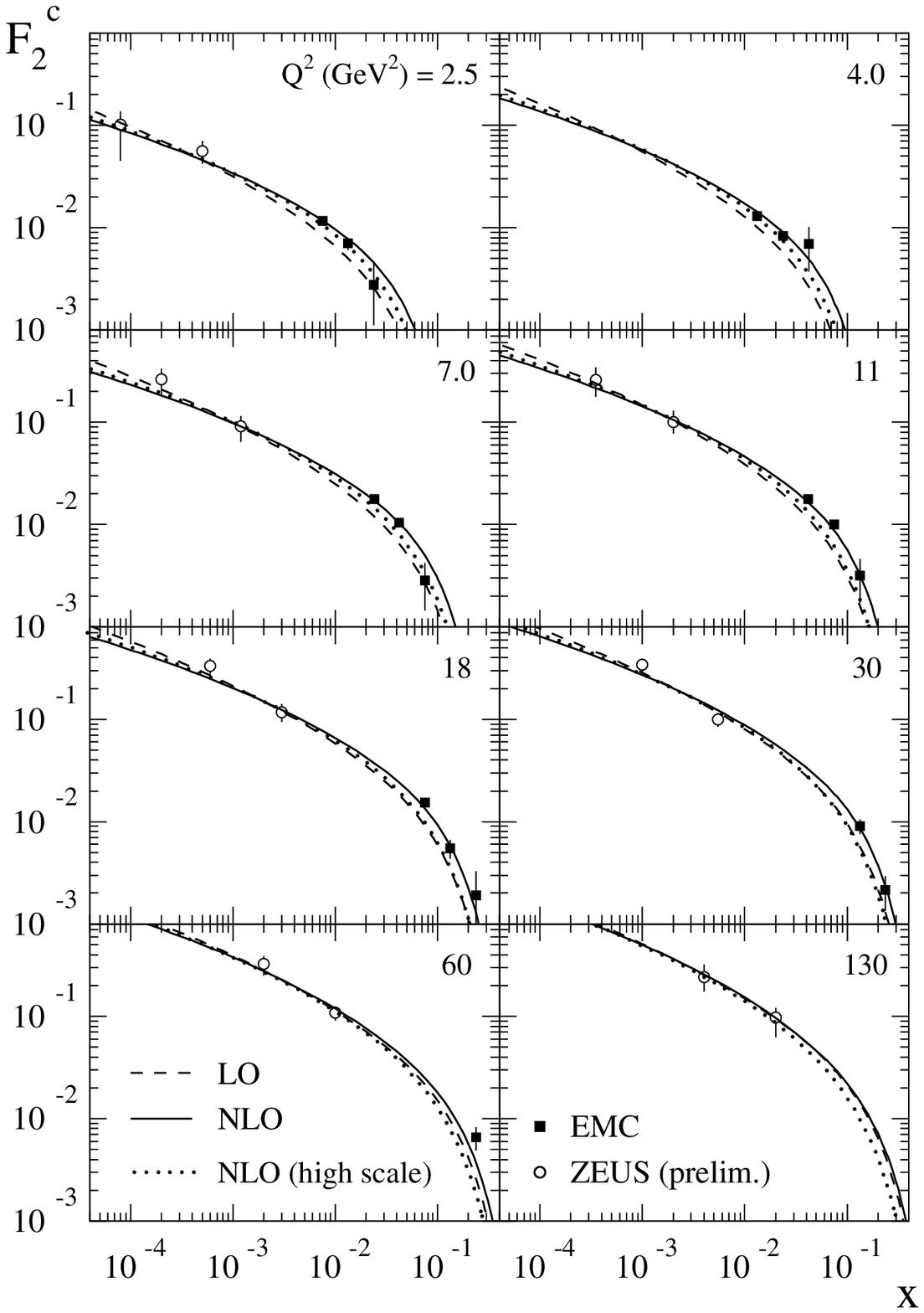,height=19.4cm}}
\vspace{3mm}
\centerline{\large\bf Fig.\ 9}
\vfill

\newpage
\vspace*{\fill}
\centerline{\hspace*{3mm}\epsfig{file=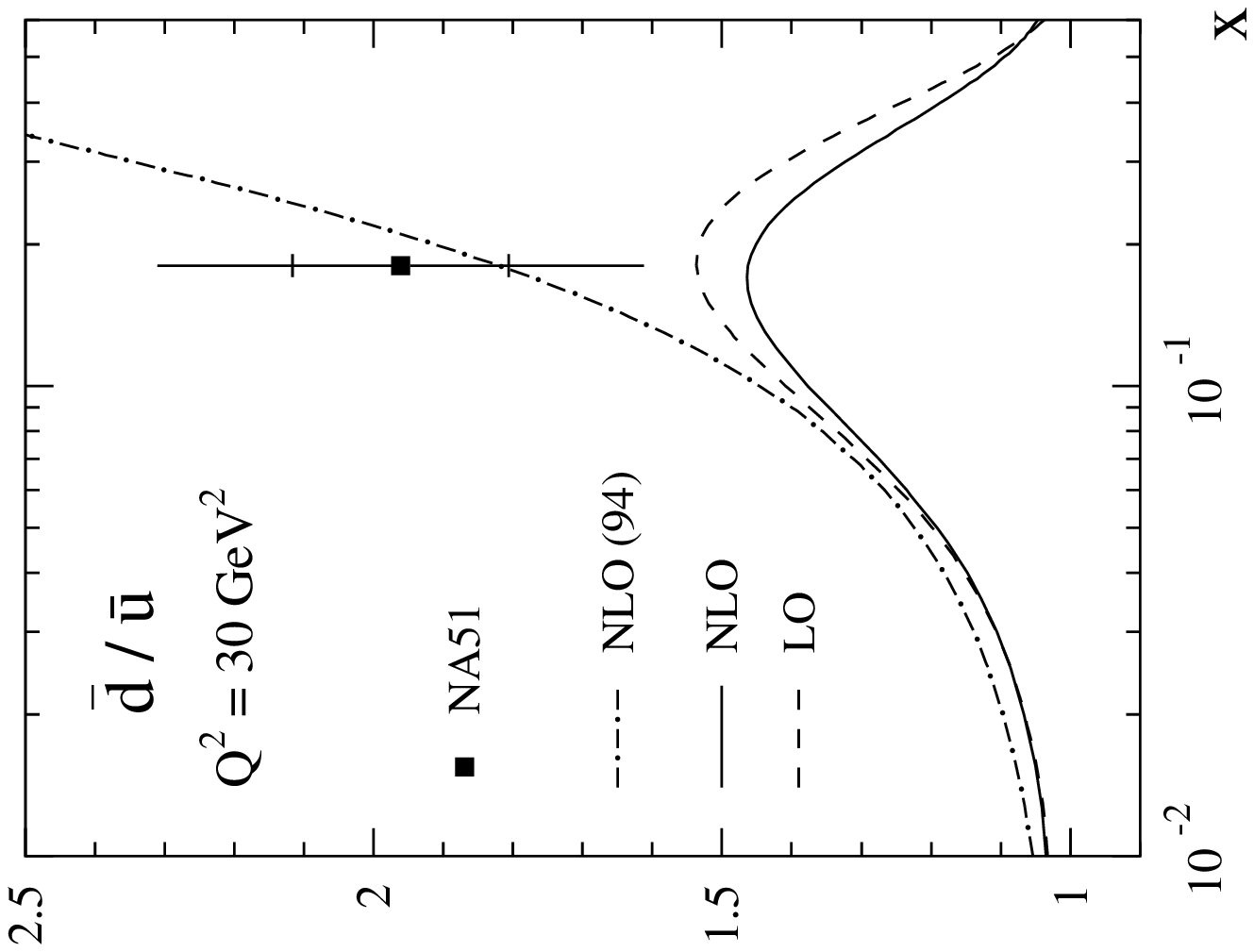,height=14cm,angle=90}}

\centerline{\hspace*{3mm}\epsfig{file=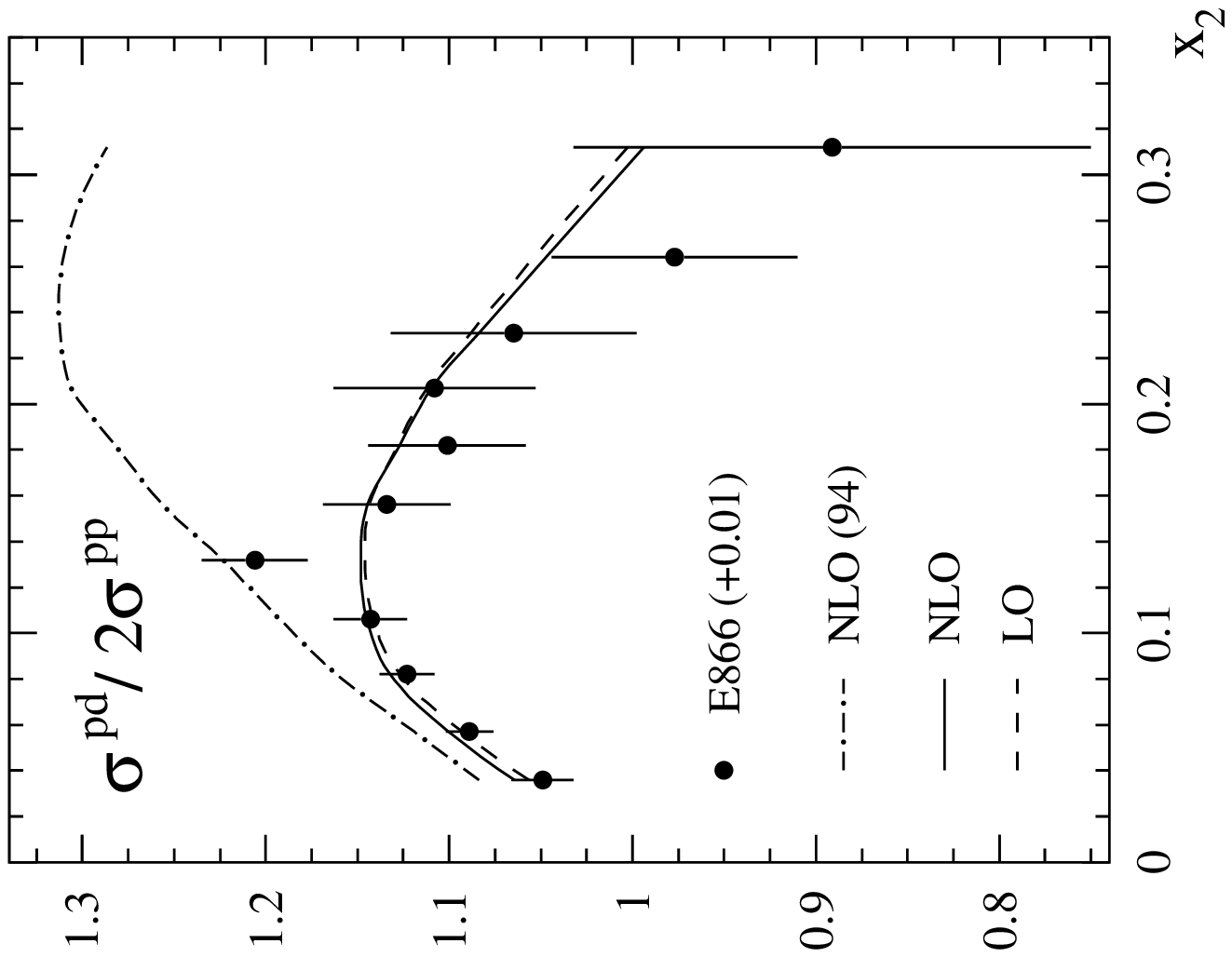,height=14cm,angle=90}}
\vspace{3mm}
\centerline{\large\bf Fig.\ 10}
\vfill

\newpage
\vspace*{\fill}
\centerline{\epsfig{file=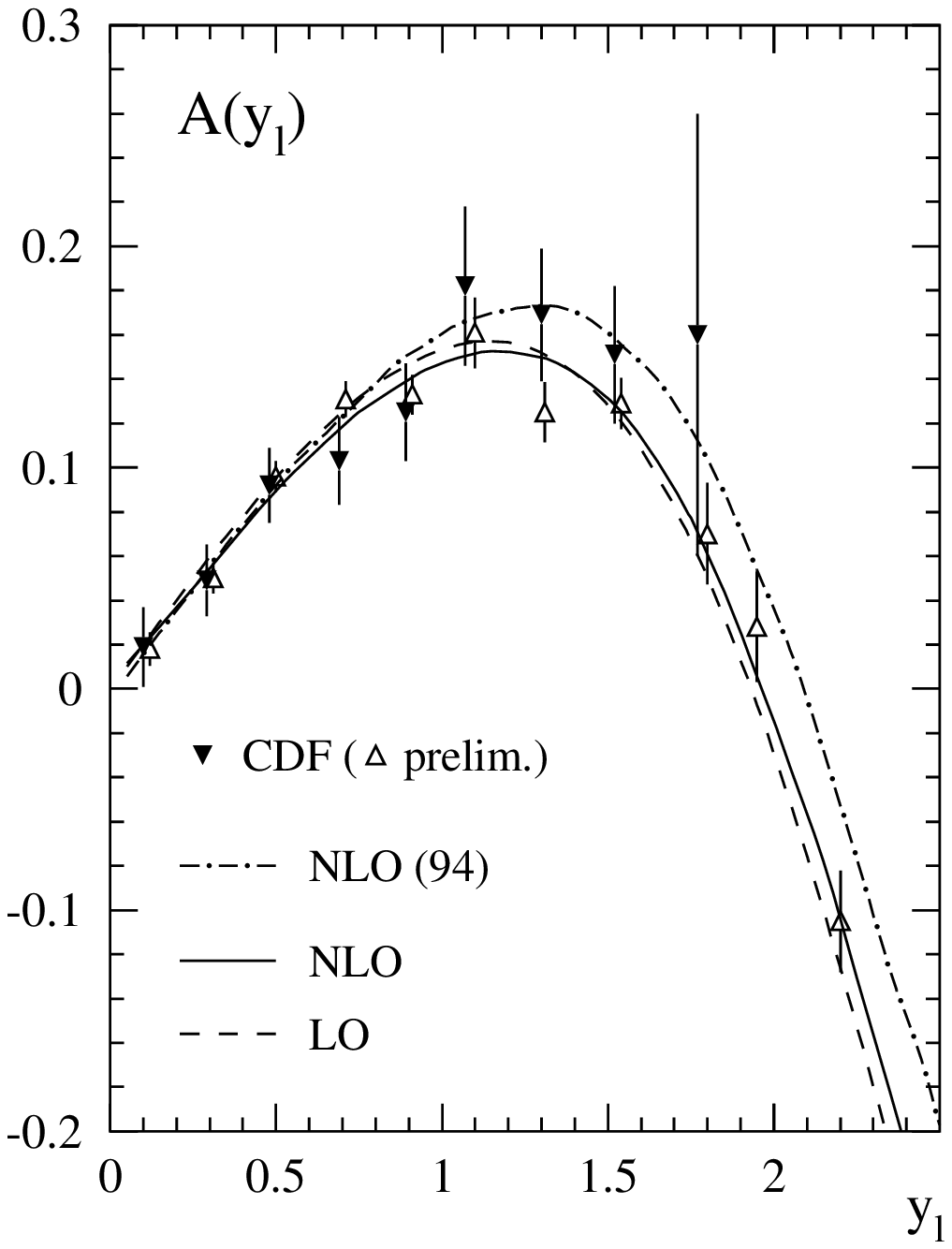,height=14cm}}
\vspace{3mm}
\centerline{\large\bf Fig.\ 11}
\vfill

\newpage
\vspace*{\fill}
\centerline{\epsfig{file=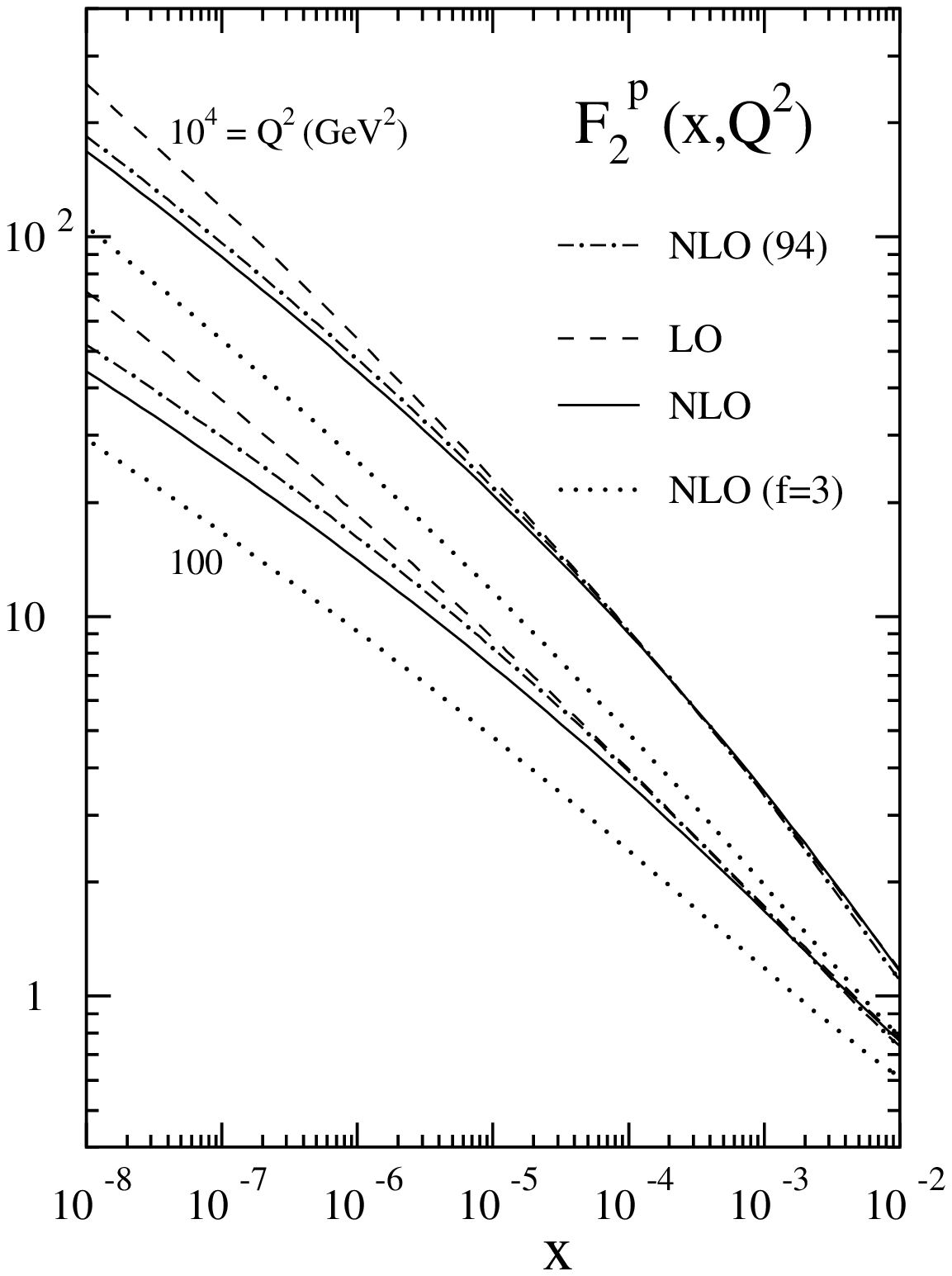,height=16.5cm}}
\vspace{3mm}
\centerline{\large\bf Fig.\ 12}
\vfill

\end{document}